\documentclass[prl,twocolumn,superscriptaddress,noshowpacs,notitlepage,longbibliography]{revtex4-1}%
\usepackage{graphicx,bm,times}
\usepackage{amsmath}
\usepackage{amsfonts}
\usepackage{amssymb}
\usepackage{color}
\usepackage{hyperref}
\usepackage{braket}
\hypersetup{
	colorlinks = true,
	allcolors = {blue}
}
\usepackage{ulem}

\newcommand{\nn}{\nonumber}
\newcommand{\kk}{\mathbf{k}}
\newcommand{\Qv}{\mathbf{Q}}

\begin{document}

	\title{Non-local $d_{xy}$ nematicity and the missing electron pocket in FeSe}
	\author{Luke C. Rhodes}
	\affiliation{School of Physics and Astronomy, University of St. Andrews, St. Andrews KY16 9SS, United Kingdom}

	\author{Jakob B\"oker}
	\affiliation{Institut f\"ur Theoretische Physik III, Ruhr-Universit\"at Bochum, D-44801 Bochum, Germany}
	
	\author{Marvin A. M\"uller}
	\affiliation{Institut f\"ur Theoretische Physik III, Ruhr-Universit\"at Bochum, D-44801 Bochum, Germany}

	\author{Matthias Eschrig}
	\email[ ]{matthias.eschrig@uni-greifswald.de}
	\affiliation{Institute of Physics, University of Greifswald, Felix-Hausdorff-Strasse 6, 17489 Greifswald, Germany}
	
	\author{Ilya M. Eremin}
	\email[corresponding author: ]{ilya.eremin@ruhr-uni-bochum.de}
	\affiliation{Institut f\"ur Theoretische Physik III, Ruhr-Universit\"at Bochum, D-44801 Bochum, Germany}

	\begin{abstract}
	The origin of spontaneous electronic nematic ordering provides important information for understanding iron-based superconductors. Here, we analyze a scenario where the $d_{xy}$ orbital strongly contributes to nematic ordering in FeSe. We show that the addition of $d_{xy}$ nematicity to a pure $d_{xz}/d_{yz}$ order provides a natural explanation for the unusual Fermi surface and correctly reproduces the strongly anisotropic momentum dependence of the superconducting gap. We predict a Lifshitz transition of an electron pocket mediated by temperature and sulphur doping, whose signatures we discuss by analysing available experimental data. We present the variation of momentum dependence of the superconducting gap upon suppression of nematicity. Our quantitatively accurate model yields the transition from tetragonal to nematic FeSe and the FeSe$_{1-x}$S$_{x}$ series, and puts strong constraints on possible nematic mechanisms.
	\end{abstract}
	\date{\today}
	\maketitle
	
	
	\section{Introduction}

	    The study of the iron-based superconductors ultimately revolves around understanding the interplay of superconductivity, magnetism and electronic nematicity. While the competition between magnetism and superconductivity has been extensively studied in other classes of superconductors, most notably the cuprates and heavy fermion compounds, the iron-based superconductors offer an ideal environment to study the interactions and consequences of the rotational symmetry breaking nematic state. 
	
	Currently, there is little consensus on the underlying mechanism that drives the formation of the electronic nematic state, with both charge-ordering and magnetic-ordering scenario's being proposed \cite{Fernandes2014}. Nevertheless, it is often assumed that the leading consequence of the nematic mechanism on the iron-based superconductors is the lifting of the degeneracy of the $d_{xz}$ and $d_{yz}$ states of the Fe atoms. This has been thought to be responsible for many of the experimentally observed changes to certain physical properties \cite{Bohmer2017}. 
	
	This idea has additionally been held for the most well studied nematic iron-based superconductor, FeSe. This material exhibits an electronic nematic state at $T_{\text{s}}=90$~K as well as a superconducting state below $T_{\text{c}}=8$~K, without the formation of magnetic ordering \cite{Bohmer2017,Kreisel2020,Shibauchi2020}. Despite a relatively small tetragonal to orthorhombic lattice distortion to this material \cite{Boehmer2013}, particularly large anisotropic effects can be observed within measurements of the resistivity anisotropy \cite{Tanatar2015}, the spin susceptibility \cite{Chen2019}, the underlying electronic structure \cite{Coldea2018} and the momentum dependence of the superconducting gap \cite{Xu2016,Sprau2017,Rhodes2018,Kushnirenko2018,Liu2018,Hashimoto2018}.
	
	However, it has become increasingly apparent that nematic order derived purely from a degeneracy breaking of the $d_{xz}$ and $d_{yz}$ states is unable to account for many of the experimental properties being observed within this material. Most notably, nematic ordering of the $d_{xz}$ and $d_{yz}$ orbitals predicts a Fermi surface topology which consists of one hole pocket and two electron pockets \cite{Mukherjee2015,Watson2016,Liang2015,Jiang2016}, whereas ARPES measurements of the nematic state of FeSe report the observation of one hole pocket and a single electron pocket \cite{Watson2017b,Yi2019,Huh2020,Cai2020,Cai2020b,Rhodes2020}. This incorrect description of the Fermi surface has also made understanding the superconducting properties of FeSe challenging \cite{Sprau2017,Benfatto2018,Rhodes2018,Hu2018}, without the addition of alternative anisotropic mechanisms, such as highly anisotropic orbital selective quasiparticle weights, \cite{Kreisel2017,Hu2018} or orbital selective spin fluctuations \cite{Benfatto2018}. 
	
	Although nematicity is often thought of in terms of a lifting of the degeneracy of the $d_{xz}$ and $d_{yz}$ states, there is one additional form of orbital ordering, involving the relevant t$_{\text{2g}}$ orbitals of the Fe atoms, which is consistent with the breaking of C$_\text{4}$ rotational symmetry. It is possible to have a non-local nematic ordering of the $d_{xy}$ states \cite{Fernandes2014b,Xing2017,Claessen2017,Christensen2020}. Although the $d_{xy}$ orbital does not conform to the B$_{\text{1g}}$ irreducible representation of the point group associated with the tetragonal space group of many iron-based superconductors, an anisotropic hopping between the $d_{xy}$ orbitals of the nearest neighbour Fe-atoms, within a 2-Fe unit cell, does. This term is typically assumed to be negligible \cite{Fernandes2014b,Christensen2020} and is often ignored within theories of the nematic state, however, there is no a-priori reason for this to be true. In fact, recent NMR \cite{Li2020} and ARPES \cite{Yi2019,Huh2020} measurements have also suggested that the $d_{xy}$ orbital may be strongly affected by the onset of the nematic state.
	
	With the goal to resolve the above mentioned inconsistencies between theory and experiment, we analyse in the following the question, what would be the consequence of nematicity, arising from the strongly correlated $d_{xy}$ orbital on the iron-based superconductors.
	
	Focusing on the case of FeSe, we show from symmetry based arguments that if a sizeable non-local nematic ordering of the $d_{xy}$ orbital is the leading form of nematic instability, then many of the discrepancies between theoretical simulations and experimental measurements can be directly resolved. In particular, the missing electron pocket at the Fermi level and the highly anisotropic momentum dependence of the superconducting gap. We justify these claims by direct comparison of our theoretical simulations of the electronic structure with existing experimental measurements. As a crucial consequence of this $d_{xy}$-dominated nematic scenario, we predict the existence of a Lifshitz transition of an electron pocket as a function of temperature and sulphur doping, which can be verified experimentally. 
	
	This result not only provides a quantitatively accurate model to study the tetragonal to nematic phase transition in FeSe, but poses strong constraints on the possible mechanism associated with the nematic instability, and ultimately, on the mechanism of superconductivity in systems where nematic order and superconducting order appear in close proximity.
	 \section{Results}   
	\subsection{The $\mathbf{B_{1g}}$ nematic state in FeSe}
          \begin{figure}[t]
	    \centering
    	\includegraphics[width = \linewidth]{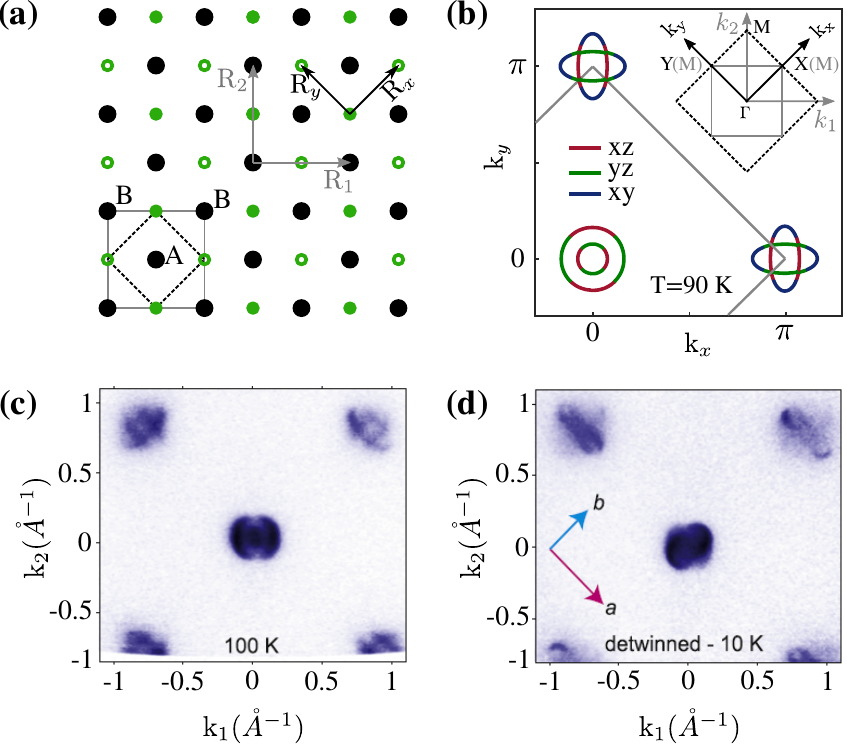}
    	\caption{\textbf{The Fermi surface of FeSe.} \textbf{a} Lattice structure of FeSe. The 2-Fe and 1-Fe unit cells of FeSe are shown as solid grey and dashed black lines with lattice vectors $(R_1,R_2)$ and $(R_x,R_y)$, respectively. Black dots denote Fe atoms. Se atoms located above (filled green) and below (empty green) the iron plane divide the system into sub-lattices A and B. \textbf{b} Fermi surface cut at $k_z = \pi$ for the tetragonal state, where the color describes the maximum orbital content of the band. The grey solid line describes the Brillouin zone boundary of the 2-Fe unit cell. Inset: Brillouin zones of the 2-Fe (solid grey) and 1-Fe (dashed black) unit cells  corresponding to the "folded" and "unfolded" Brilloin zone, respectively. \textbf{c} Fermi surface map measured by ARPES of the tetragonal phase of FeSe at 100~K and \textbf{d} of the detwinned sample in the orthorhombic phase at 10~K taken from Ref.~\cite{Watson2017b} under the Creative Commons Attribution 4.0 International License.}  
    	\label{Fig:1}
        \end{figure}
        \begin{table*}
        	\includegraphics[width = \linewidth]{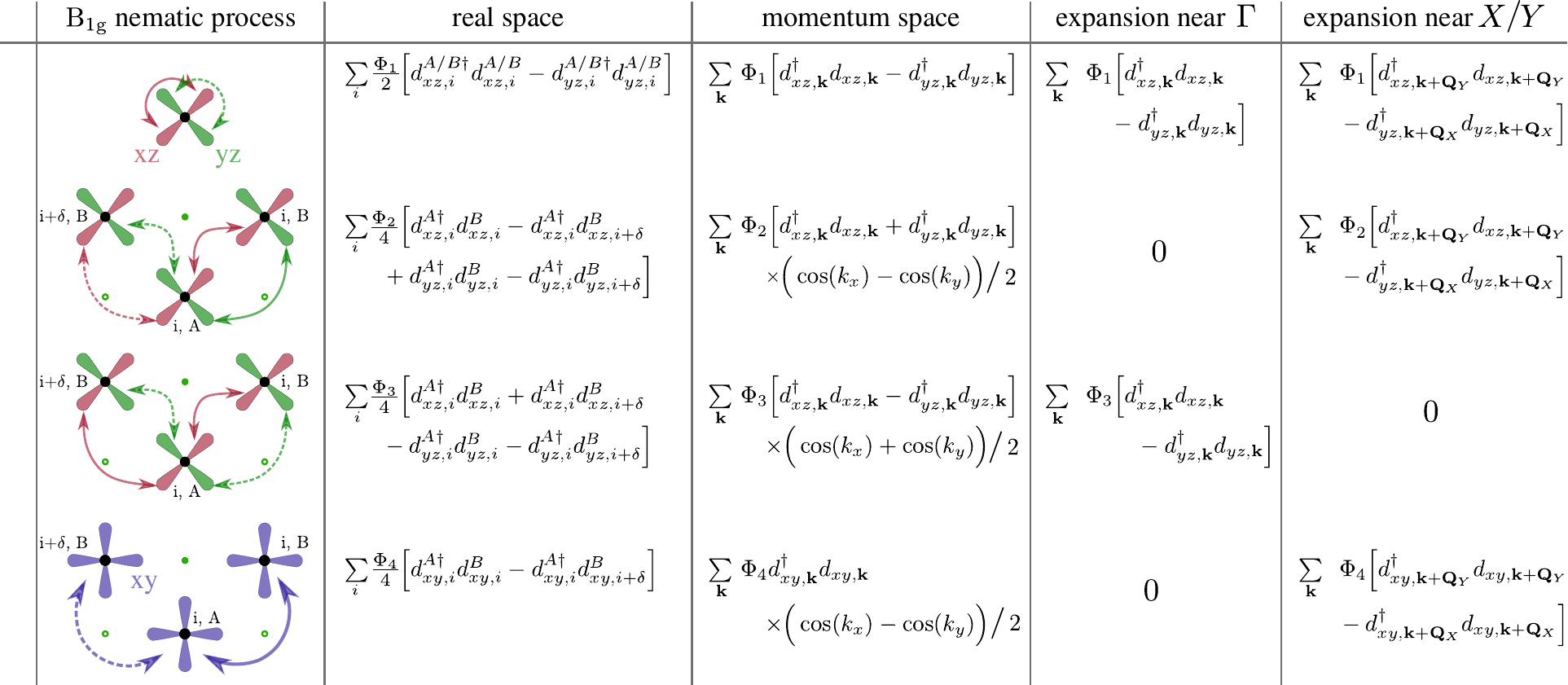}
    \caption{\textbf{All B$_{\text{1g}}$ nematic order parameters for FeSe.} All possible nematic order parameters within the B$_{\text{1g}}$ irreducible representation of the P4/nmm space group involving interactions between $d_{xz}$, $d_{yz}$ and $d_{xy}$ orbitals within a single 2-Fe unit cell. Results presented in real and momentum space as well as Taylor-expanded in the vicinity of the high symmetry points (following the notations of Ref.~\cite{Fernandes2014b}). Here we use coordinates of the unfolded Brilloiun zone. A sketch of of each nematic process is depicted in the first column where solid (dashed) arrows are associated with a positive (negative) amplitudes of the corresponding nematic order parameter. Note that $d_{xy}$ nematicity is a purely non-local phenomenon.}
\label{Tab_NematicStates}
        \end{table*}
To understand the consequence of non-local $d_{xy}$ nematicity on the nematic state of FeSe we begin with a pedagogical discussion on the B$_{\text{1g}}$ symmetric order parameters of the D$_{\text{4h}}$ point group of the P4/nmm space group on the electronic structure of FeSe. 
	
	Within the P4/nmm crystal structure, the Fe atoms lie in-plane on a square grid, and are connected by staggered out-of-plane pnictide or chalcogenide atoms, which gives rise to a two-Fe atom unit cell, as shown in Fig.~\ref{Fig:1}(a). This 2-Fe unit cell has primitive in-plane lattice vectors $R_1 = R_x-R_y$ and $R_2=R_x+R_y$ which generate a  Brillouin zone, defined by the coordinates $k_1 = k_x - k_y$, $k_2 = k_x + k_y$ and $k_z$, as shown in the inset of Fig.~\ref{Fig:1}(b). In this 2-Fe Brillouin zone, the important high symmetry points are the $k_z = 0$, $\Gamma$ point and the $k_z=\pi$, $Z$ point at the zone center $(k_1,k_2,k_z) =(0,0,k_z)$ and the $k_z = 0$, $M$ point and $k_z =\pi$ $A$ point at the zone corner $(k_1,k_2,k_z) =  (\pi,\pi,k_z)$.  It is, however, useful to discuss the electronic structure of the P4/nmm lattice in terms of the lattice vectors $R_x$ and $R_y$, and the momentum $k_x$ and $k_y$, particularly to illustrate the effect of $C_{\text{4}}$ symmetry breaking. In this case, the Brillouin zone can be “unfolded” into an 45 degree rotated effective 1-Fe Brillouin zone, where the states around the $M$ and $A$ point are mapped to the $X$ point at $(k_x,k_y,k_z) = (\pi,0,k_z)$ and the $Y$ point at $(k_x,k_y,k_z) = (0,\pi,k_z)$. Here, our model will retain the physical two-atom unit cell periodicity, however we will discuss symmetry breaking in terms of the coordinates $k_x$ and $k_y$. 
	
	As a starting point, we employ the low-energy model of the $d_{xz}$, $d_{yz}$ and $d_{xy}$ orbitals of the Fe atoms, following Ref.~\cite{Cvetkovic13}, and fit the parameters to quantitatively agree with ARPES data of the tetragonal state electronic structure of FeSe near the $\Gamma$ ($Z$) point and $M$ ($A$) point respectively \cite{Watson2016,Coldea2018,Rhodes2017},
	\begin{align}
	H_{\Gamma/\text{Z}} = H_{0}^{\Gamma/Z} + H_\text{SOC}^{\Gamma/Z} + H_{\text{nem}}^{\Gamma/Z},
	\end{align}
	and 
	\begin{align}
	H_{M/A} = H_{0}^{M/A} + H_\text{SOC}^{M/A} + H_{\text{nem}}^{M/A}.
	\end{align}
	Here the kinetic part $H_{0}^{\Gamma/Z} (H_{0}^{M/A} )$ and the atomic on site spin-orbit coupling $H_\text{SOC}^{\Gamma/Z} (H_\text{SOC}^{M/A}) $ are given explicitly in the supplementary material \cite{Supp}. The possible nematic Hamiltonian at the Brillouin zone center $H_{\text{nem}}^{\Gamma/Z}$ and at the Brillouin zone corner $H_{\text{nem}}^{M/A}$ are discussed below. 
	
	For the tetragonal state, $H_{\text{nem}}^{\Gamma/Z} = H_{\text{nem}}^{M/A} = 0$, this model describes a quasi-2D electronic structure which exhibits two corrugated $d_{xz}/d_{yz}$ hole pockets at the Brillouin zone center with the inner hole pocket only present near $k_z = \pi$, and two almost cylindrical $d_{xz}/d_{xy}$ and $d_{yz}/d_{xy}$ electron pockets at the Brillouin zone corners. The Fermi surface for this model is presented in Fig. \ref{Fig:1}(b). The states at the zone corner have majority $d_{xz}$ ($d_{yz}$) character along the length of the ellipse, but $d_{xy}$ at the outer tips. This model is in direct agreement with the low energy tetragonal electronic structure measured by ARPES, the Fermi surface of which, from Ref.~\cite{Watson2017b}, is presented in Fig.~\ref{Fig:1}(c). In Fig.~\ref{Fig:1}(d) we also show the experimentally measured Fermi surface of the nematic state of FeSe, which the nematic order parameter must ultimately reproduce.
	
	If we focus on the $d_{xz}$, $d_{yz}$ and $d_{xy}$ orbitals of the Fe atoms, within a single two atom unit cell, then we can write down four possible C$_\text{4}$ symmetry breaking nematic orders in the B$_{\text{1g}}$-channel, which arise either from on-site interactions on the Fe-atoms or from interactions between two neighbouring Fe-atoms on different sublattices. We list these four order parameters in Table 1 along with a schematic cartoon of the nematic process, mathematical formulation in real and momentum space and the effect each term has on the hole and electron pockets. 
	
\begin{figure}
    \centering
    \includegraphics{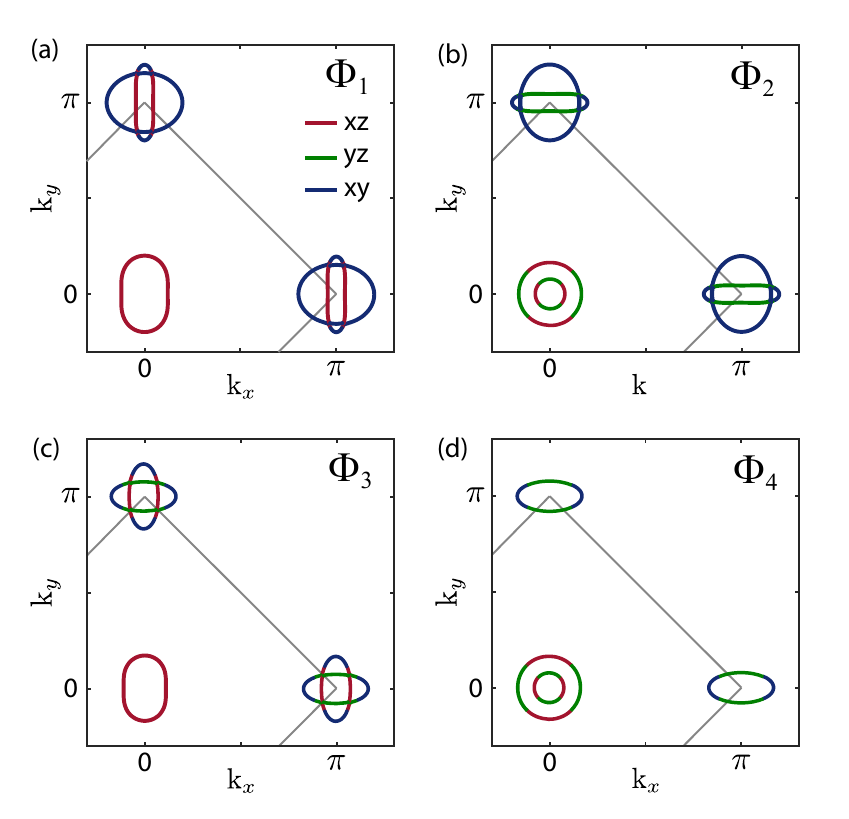}
    \caption{\textbf{Individual consequence of the nematic order parameters on the tetragonal state Fermi surface of FeSe}. \textbf{a} $\Phi_1$=26~meV. \textbf{b} $\Phi_2$ = -26~meV. \textbf{c} $\Phi_3$=15~meV. \textbf{d} $\Phi_4>$50~meV.}
    \label{Fig:2}
\end{figure}

	The first term, $\Phi_1$, describes a degeneracy breaking of the on-site $d_{xz}$ and $d_{yz}$ states. This form, often referred to as ferro-orbital ordering \cite{Lee2009}, generates an equivalent splitting of the $d_{xz}$ and $d_{yz}$ bands which affects both the hole and electron pockets equally, as shown in Fig.~\ref{Fig:2}(a). This type of nematic order was originally discussed in the literature \cite{Mukherjee2015,Suzuki2015,Nakayama2014,Watson2015}, however, it later became apparent that whilst the evolution of the hole pocket is correctly captured by $\Phi_1$, the evolution of the electron pocket into a $d_{xz}$ dominated peanut, elongated along the $k_y$ axis, and a larger outer $d_{xy}$ dominated ellipse was not consistent with experiment \cite{Watson2016,Suzuki2015,Pfau2019,Fanfarillo2016,Fedorov2016,Zhang2015}. It was realised that a sign change in the nematic order parameter between the hole and electron pocket, involving the $d_{xz}$ and $d_{yz}$ orbitals was required \cite{Fanfarillo2016,Fedorov2016,Watson2016}. Interestingly, this sign change of the nematic order between electron and hole pockets is analogous to the sign-changing $s^{\pm}$-wave superconducting gap and indicates that the same electronic interactions are most likely in play \cite{Xing2017,Claessen2017}.      
	
	The second and third terms, $\Phi_2$ and $\Phi_3$, describe non-local nematic ordering between nearest neighbour atoms involving the $d_{xz}$ and $d_{yz}$ states. $\Phi_2$ describes an anisotropic hopping potential such that hopping is favoured in the $x$ direction, but disfavoured in the $y$ direction for both orbitals. This has been referred to as d-wave nematic order within the literature \cite{Jiang2016,Kang2018,Kreisel2015,Liang2015,Fanfarillo2016}. Conversely, $\Phi_3$ describes a global increase in the hopping of the $d_{xz}$ orbital, and a global decrease in the hopping of the $d_{yz}$ orbital, often referred to as extended s-wave nematic order \cite{Jiang2016,Liang2015,Watson2016}. As shown in Table 1,  $\Phi_2$ only affects the electron pockets at the $X/Y$ point, and $\Phi_3$ only affects the hole pocket at the $\Gamma$ point. We present the Fermi surface generated by the inclusion of $\Phi_2$ or $\Phi_3$ in Fig. \ref{Fig:2}(b) and \ref{Fig:2}(c) respectively.  
	
	The fourth term, $\Phi_4$ is a purely non-local nematic ordering between the $d_{xy}$ states of nearest neighbour atoms, analogous to that of $\Phi_2$. This term, has been mentioned as a previously possible form of nematic order \cite{Xing2017,Fernandes2014b}, however it was assumed to only produce a minor effect on the electronic band dispersion \cite{Kang2018,Christensen2020} and is often neglected entirely. If however, $\Phi_4$ is made sufficiently large ($> 50$~meV) then it would be possible to generate a Lifshitz transition, which can induce a one-electron pocket Fermi surface. We show the generated Fermi surface obtained from the inclusion of $\Phi_4 > 50$~meV in Fig. \ref{Fig:2}(d). 
As mentioned above, these are the only four symmetry allowed order parameters that can describe the evolution of the nematic state within the P4/nmm space group. Therefore, some combination of these four order parameters should be able to reproduce the Fermi surface of the nematic state of FeSe as measured by ARPES. As shown in Table 1, $\Phi_1$ contributes equally to both the states around the $\Gamma$ point and the $X/Y$ point. However, $\Phi_2$ and $\Phi_3$ contribute solely to the $X/Y$ and $\Gamma$ point, respectively. It is therefore not possible to fit the magnitude of $\Phi_1$, $\Phi_2$ and $\Phi_3$ directly to experiment. This is because experimental measurements can only extract the total change in the band dispersion, which for the hole pocket is defined as  $\Phi^{h}_{xz,yz} = \Phi_1+\Phi_3$ and for the electron pocket is $\Phi^{e}_{xz,yz} = \Phi_1$ + $\Phi_2$. In this model we therefore write the nematic order parameters as
	\begin{align}
	H_{\text{nem}}^{\Gamma/Z} =\sum_{\kk\sigma}\Phi^{h}_{xz,yz}(d^\dagger_{xz,\kk\sigma}d_{xz,\kk\sigma}-d^\dagger_{yz,\kk\sigma}d_{yz,\kk\sigma})
	\label{Eq:Gamma_NematicOrder}
	\end{align}
	for the hole pocket, and 
	\begin{align}
	H^{A/M}_{\text{nem}}= \sum_{\kk} \Big[\Phi^{e}_{xz,yz}&(d^\dagger_{xz,\kk+\textbf{Q}_Y\sigma}d_{xz,\kk+\textbf{Q}_Y\sigma}\nonumber \\&-d^\dagger_{yz,\kk+\textbf{Q}_X\sigma}d_{yz,\kk+\textbf{Q}_X\sigma})\nonumber \\
	+\Phi_{4}&(d^\dagger_{xy,\kk+\textbf{Q}_Y\sigma}d_{xy,\kk+\textbf{Q}_Y\sigma}\nonumber \\ &-d^\dagger_{xy,\kk+\textbf{Q}_X\sigma}d_{xy,\kk+\textbf{Q}_X\sigma})\Big].
	\end{align}
	for the electron pocket.
	We then fit the parameters $\Phi^{h}_{xz,yz}$, $\Phi^{e}_{xz,yz}$ and $\Phi_4$ directly to experimental values \cite{Watson2017b,Yi2019,Huh2020}. 
\begin{figure*}[t]
	\centering
	\includegraphics[width=0.7\linewidth]{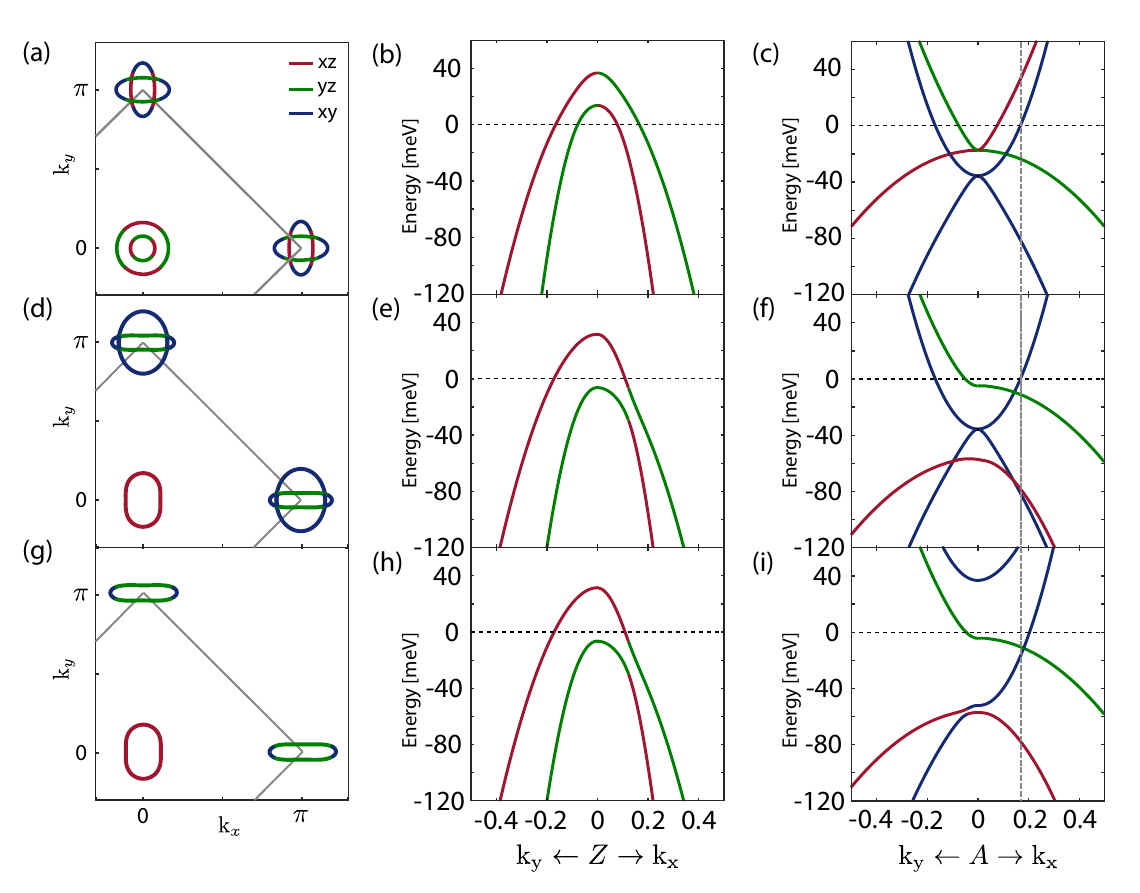}
	\caption{\textbf{Effect of non-local $d_{xy}$ nematicity}. Comparison of the Fermi surface and band dispersions around the $Z$, $(0,0,\pi)$, point and $A$, $(\pi,0,\pi)$, point for \textbf{a-c} The tetragonal electronic structure, with $\Phi^{h}_{xz,yz}$=$\Phi^{e}_{xz,yz}$ =$\Phi_4$=0. \textbf{d-f} A purely $d_{xz}/d_{yz}$ nematic scenario where $\Phi^{h}_{xz,yz}$=15~meV, $\Phi^{e}_{xz,yz}$ = -26~meV and $\Phi_4$=0. \textbf{g-i} A scenario with additional non-local $d_{xy}$ nematicity and Hartree shift $\Phi^{h}_{xz,yz}$=15~meV, $\Phi^{e}_{xz,yz}$ = -26~meV, $\Phi_4$=45~meV and $\Delta\epsilon_{xy}$=40~meV.}
	\label{Fig:3}
\end{figure*}
In Fig 3(a-c), we present the Fermi surface and band dispersion around the $\Gamma$ and $M$ point for the tetragonal state, i.e $\Phi^{h}_{xz,yz}$=$\Phi^{e}_{xz,yz}$ =$\Phi_4$=0. In Fig, 3(d-f) we next present the Fermi surface and band dispersion's using $\Phi^{h}_{xz,yz}$=15~meV, $\Phi^{e}_{xz,yz}$ = -26~meV and keep $\Phi_4$ = 0~meV. This purely $d_{xz}/d_{yz}$ nematic order correctly describes the 37~meV band separation of the two hole bands (which is obtained by a combination of the nematic splitting and spin orbit splitting where the spin orbit coupling parameter, $\lambda^{h}_{SOC}=23$~meV \cite{Supp}) and correctly describes the shape and orbital character of the $d_{xz}$ dominated elliptical hole pocket. Additionally, the experimentally observed peanut shaped electron pocket with $d_{yz}$ orbital content along the length of the peanut, and $d_{xy}$ orbital content at the tips, is also quantitatively captured. However, this scenario, with $\Phi_4$ = 0~meV, also describes a second large electron pocket of predominantly $d_{xy}$ orbital character, which is not detected in experimental measurements. This type of order parameter is the main form considered within the literature \cite{Kreisel2020,Yu2020_arXiv,Coldea2018}.
	
In Fig. 3(g-i) we now present our scenario, where we use the same values of $\Phi^{h}_{xz,yz}$ and $\Phi^{e}_{xz,yz}$ as in Fig. 3(d-f) but now we set $\Phi_4 = 45$~meV. We note that the $\Phi_4$ term, as it is discussed in Table~\ref{Tab_NematicStates}, lifts the degeneracy of the $d_{xy}$ saddle point located at -50~meV at the $M$ ($A$) point in the tetragonal state. This term alone however would ensure that the splitting of the saddle point is symmetric, such that the lower branch of the $d_{xy}$ band would also be pushed to a lower binding energy. Since a large shift in the lower $d_{xy}$ band is not observed in the experimental data \cite{Coldea2018} we add a finite energy shift to the on-site $d_{xy}$ energy in the nematic state
	\begin{align}
	H_0^{M/A} \rightarrow 
	H_0^{M/A} + \sum_{\kk,\sigma} \Delta\epsilon_{xy}(&d^\dagger_{xy,\kk+\textbf{Q}_Y\sigma}d_{xy,\kk+\textbf{Q}_Y\sigma}\nonumber\\&+d^\dagger_{xy,\kk+\textbf{Q}_X\sigma}d_{xy,\kk+\textbf{Q}_X\sigma})
	\label{Eq:HartreeTerm}
	\end{align}
	which then generates an asymmetric nematic splitting. This term has the consequence that the lower branch of the $d_{xy}$ saddle point, remains in place, whilst the upper branch increases in energy by $\Phi_4 + \Delta\epsilon_{xy}$ where a value of $\Delta\epsilon_{xy} =40$~meV was found to best fit the experimental data. The result of this, is that the upper $d_{xy}$ band is placed 36~meV above the Fermi level as shown in Fig. 3(i). Note that in general the onsite energy $\epsilon_{xy}$ can change when entering the nematic state which corresponds to a Hartree shift of the $d_{xy}$ states. While we do not explicitly calculate this Hartree shifts of the orbitals self-consistently, we do calculate the renormalization of the total chemical potential when entering into the nematic phase and show that it is indeed strongly affected by the transition.  We explicitly assume that this Hartree shift has the same mean-field-like temperature dependence as the nematic order. It is important to note that this shift of the onsite energy only affects the precise size of the peanut shaped electron pocket and is not needed to obtain the correct Fermi surface topology of FeSe, which may be obtained if we set $\Phi_4$ larger than 50~meV. It is however important to yield the correct position of the $d_{xy}$ dominated bands far below the Fermi energy, which is our reason for including it here. Note, as indicated by the vertical dashed line in Fig.~\ref{Fig:3}(c,f,i), we find that the Fermi momentum ($\kk_{\text{F}}$) associated with $d_{xy}$ states of the peanut tips to be smaller in the tetragonal phase (c) than in the orthorhombic phase for our parametrization in (i) which is in good agreement with Ref.~\cite{Yi2019,Watson2017b}. This is not the case if one neglects the $d_{xy}$ nematic order as seen in (f).
	
	Finally, we point out that the exact values for $\Phi_{4}$ and $\Delta\epsilon_{xy}$ are not particularly well constrained as ARPES data can not detect the actual position of the upper $d_{xy}$ band. However, as we will discuss below, a value of roughly this magnitude is required to correctly describe experimental properties as a function of temperature and sulphur doping.
	
	The importance of a sizeable nematic order between $d_{xy}$ orbitals in the correct description of the electronic structure of FeSe is the main feature of our model of nematicty in FeSe. In contrast to the previous proposals \cite{Christensen2020,Kreisel2018,Yu2020_arXiv} this scenario does not require strong incoherence of the $d_{xy}$-dominated electron pocket. Furthermore, as we also discuss below, this nematic order between $d_{xy}$ orbitals is also required to resolve the multiple discrepancies between theory and experiment discussed in the literature \cite{Rhodes2018,Rhodes2019,Benfatto2018,Kreisel2018,Hu2018}.
\begin{figure*}[t]
	\centering
	\includegraphics[width=1\linewidth]{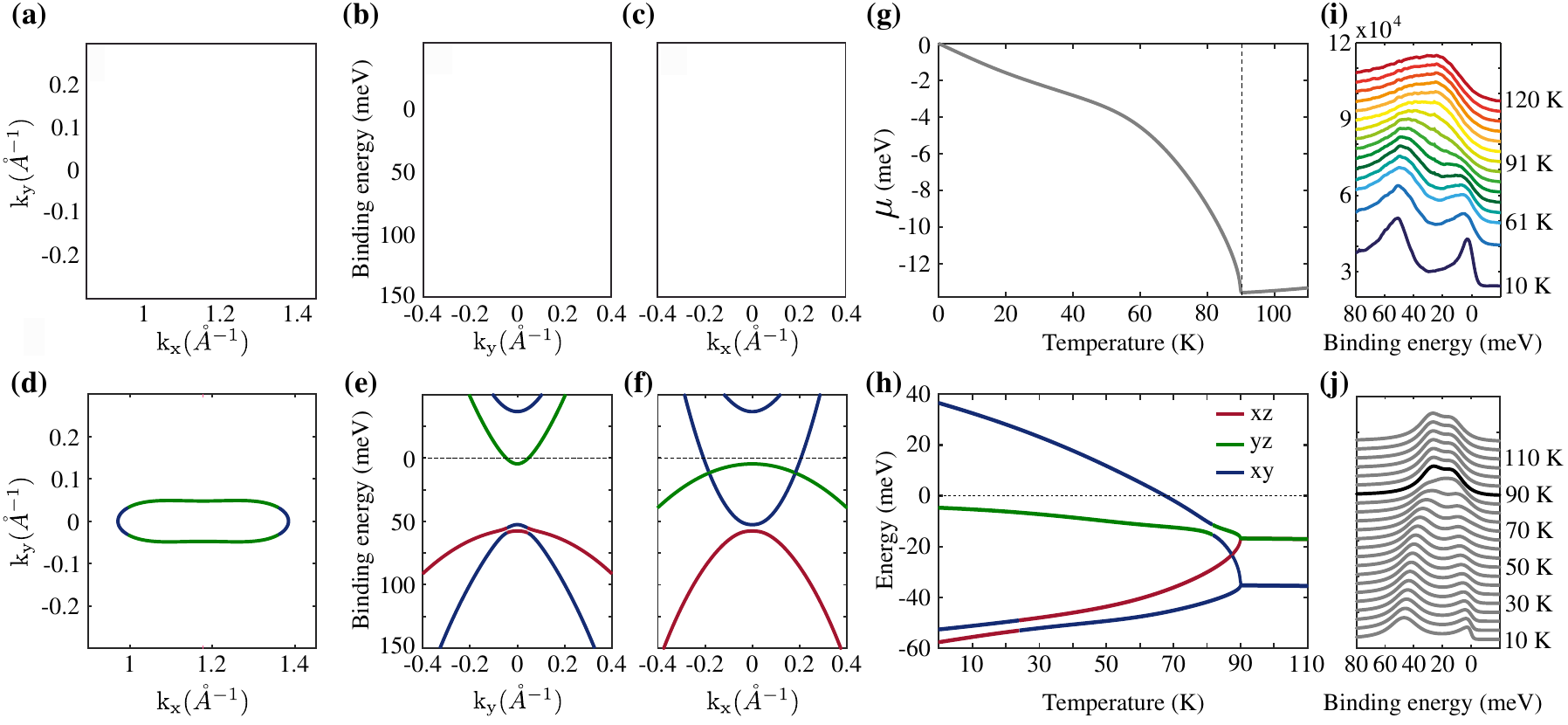}
	\caption{\textbf{Comparison with ARPES data.} \textbf{a} Experimental Fermi surface of the $A$ point of FeSe measured by ARPES on a detwinned crystal taken from Ref.~\cite{Watson2017b}. \textbf{b,c} Experimental band dispersion along the $k_x$ and $k_y$ axis respectively centered around the $A$ point taken from Ref.~\cite{Watson2017b}. \textbf{d-f} Corresponding simulated Fermi surface topology and band dispersions of the orbital projective model with the nematic order parameter. \textbf{g} Change in the chemical potential as a function of temperature using a mean field temperature dependence $\Phi_i(T) = \Phi_i(0)\sqrt{1-(\frac{T}{T_{\text{s}}})}$. \textbf{h} Corresponding change to the band positions at the $A$ point as a function of temperature. \textbf{i} Experimental temperature dependence of the spectral function taken from Ref.~\cite{Watson2016}. \textbf{j} Simulated spectral function as a function of temperature calculated using Eq.~\eqref{Eq:SpectralFunction} with a broadening parameter of $\Gamma\sim10$ meV.}
	\label{Fig:4}
\end{figure*}
\subsection{Experimental evidence for non-local $d_{xy}$ nematicity}
The main consequence of a sizeable non-local $d_{xy}$ nematic term is the generation of a Lifshitz transition at the corner of the Brillouin zone shortly after the onset of the nematic state. This Lifshitz transition is directly responsible for the evolution between a two-electron pocket Fermi surface in the tetragonal state into a one-electron pocket Fermi surface in the nematic state. Here we discuss available experimental data which can be interpreted in terms of this Lifshitz transition driven by nematic order and compare these experiments to our model calculations. \\ In Fig.~\ref{Fig:4} we now compare our simulated electronic structure of the $A$ point of FeSe, using the parameters discussed in Fig. \ref{Fig:3}(g-i), with the experimental ARPES data of detwinned crystals of FeSe from Ref.~\cite{Watson2017b}. In Fig.~\ref{Fig:4}(d) we present a close up of the Fermi surface predicted around the $A$ point of our model which consists of a single peanut shaped pocket with dominant $d_{yz}$ orbital character along the length of the peanut and dominant $d_{xy}$ orbital character at the tips. The shape and size of this pocket is in direct agreement with the experimental measurement shown in  Fig.~\ref{Fig:4}(a) as well as the data published in Ref.~\cite{Yi2019,Huh2020,Rhodes2020}. In Fig.~\ref{Fig:4}~(e,f) we present a cut of the band dispersion centred at the $A$ point along the $k_x$ and $k_y$ axis, respectively. Our model exhibits a $d_{yz}$ saddle point located at -5~meV below $E_{\text{F}}$, which can be observed along the $k_x$ axis as a hole like dispersion that almost reaches $E_{\text{F}}$ and along the $k_y$ axis as a very shallow electron like dispersion. A second saddle point of $d_{xy}$ orbital content is also present at -55~meV, which can equivalently be observed as an electron like dispersion along $k_x$ and a hole like dispersion along $k_y$. Finally, a hole-like $d_{xz}$ band is also located around -60~meV along both the $k_x$ and $k_y$ axis. This simulated band dispersion is entirely consistent with the ARPES data presented in Fig.~\ref{Fig:4}(b,c), as well as the data published on detwinned crystals of FeSe in Ref.~\cite{Yi2019,Huh2020,Rhodes2020}. However in the experimental data the $d_{xy}$ and $d_{xz}$ bands at $\sim$-50~meV can not be separately resolved due to self energy broadening effects \cite{Aichhorn2010,Acharya2020_arXiv}. This electronic structure is also fully consistent with ARPES measurements on "twinned" crystals, \cite{Watson2015,Fedorov2016,Fanfarillo2016,Kushnirenko2017,Watson2016,Kushnirenko2018,Rhodes2018}.
	
	Despite the removal of an entire electron pocket within the nematic state, we find that the Luttinger theorem is conserved within this model. In Fig.~\ref{Fig:4}(g) we plot the change in chemical potential $\mu(T)$ self consistently calculated at a fixed number of particles and using a mean field temperature dependence of the nematic order parameter of the form $\Phi_i(T) = \Phi_i(0)\sqrt{1-(\frac{T}{T_{\text{s}}})}$, see Ref.~\cite{Supp} for details.  We find that, within our parametrization, a shift of 13~meV is required to ensure particle number conservation between the tetragonal and nematic state, which is in good agreement with the 10~meV shift extracted from ARPES measurements \cite{Watson2017,Watson2015}. Note that our parametrization accounts for charge neutrality, which is conserved throughout the whole doping range by the renormalization of the chemical potential $\mu$. We emphasize that charge neutrality is also conserved at the Lifshitz transition of the $d_{xy}$ dominated electron pocket.  Additionally, in Fig.~\ref{Fig:4}(h) we present the evolution of the band positions at the $A$ point calculated using the same temperature dependence and chemical potential shift.  Under this assumption, we see that the Lifshitz transition occurs at $T_{\text{LT}}\sim$70~K, shortly after the nematic transition at $T_{\text{s}} =90$~K.
	
	Finally, in Fig.~\ref{Fig:4}(j) we plot the temperature dependence of the spectral function,
	\begin{equation}
	A(\kk,\omega) = -\frac{1}{\pi} \text{Tr}\{ \text{Im}{G(\kk,\omega)} \}f(\omega,T),
	\label{Eq:SpectralFunction}
	\end{equation} 
	at the $A$ point, where $G(\kk,\omega) = [\omega - (H(\kk)-\mu) + i\Gamma ]^{-1}$ and $f(\omega,T)$ is the Fermi function. 
	Here we have used an energy broadening parameter of $\Gamma = 10$~meV for all bands to simulate self energy broadening. We see that between $0<T<70$~K the $d_{xy}$ band below the Fermi level resides very close in energy to the $d_{xz}$ band. Consequently, in the spectral function these two bands merge to a single broad peak at low temperatures, which together with the $d_{yz}$ excitation near $E_{\text{F}}$ suggests that two peaks would be observed by the spectral function, which is in good agreement with the experimental measurements reported in Ref.
	\cite{Watson2016,Fanfarillo2016,Yi2019,Fedorov2016}. We show the experimental measurement of Ref.~\cite{Watson2016}, in Fig.~\ref{Fig:4}(i) for comparison.
	
	We note that there has been some ambiguity as to the experimental interpretation of this spectral function, particularly regarding the orbital content of the lower peak at -50~meV. With some authors claiming that the lack of a clearly observable splitting of the peak closest to $E_{\text{F}}$ in the spectral function at 100~K indicates that the $d_{xz}$ and $d_{yz}$ states must remain degenerate \cite{Watson2016,Fedorov2016,Rhodes2018,Watson2017}. However, in our simulation, we find that the direct observation of this splitting would be challenging, particularly with large self energy broadening effects that are present at the $A$ point in the tetragonal state \cite{Watson2016,Fedorov2016,Fanfarillo2016,Acharya2020_arXiv}. Coupled with the observation of $d_{xz}$ spectral weight of the band at -50~meV \cite{Fanfarillo2016,Yi2019,Huh2020}, as well as the symmetry  allowed possibility of an orbital transmutation of $d_{xz}$ and $d_{xy}$ weight as a function of temperature \cite{Christensen2020}, we find that our interpretation is entirely consistent with the measured ARPES data. 
   		\begin{figure}
			\centering
			\includegraphics[width=0.8\linewidth]{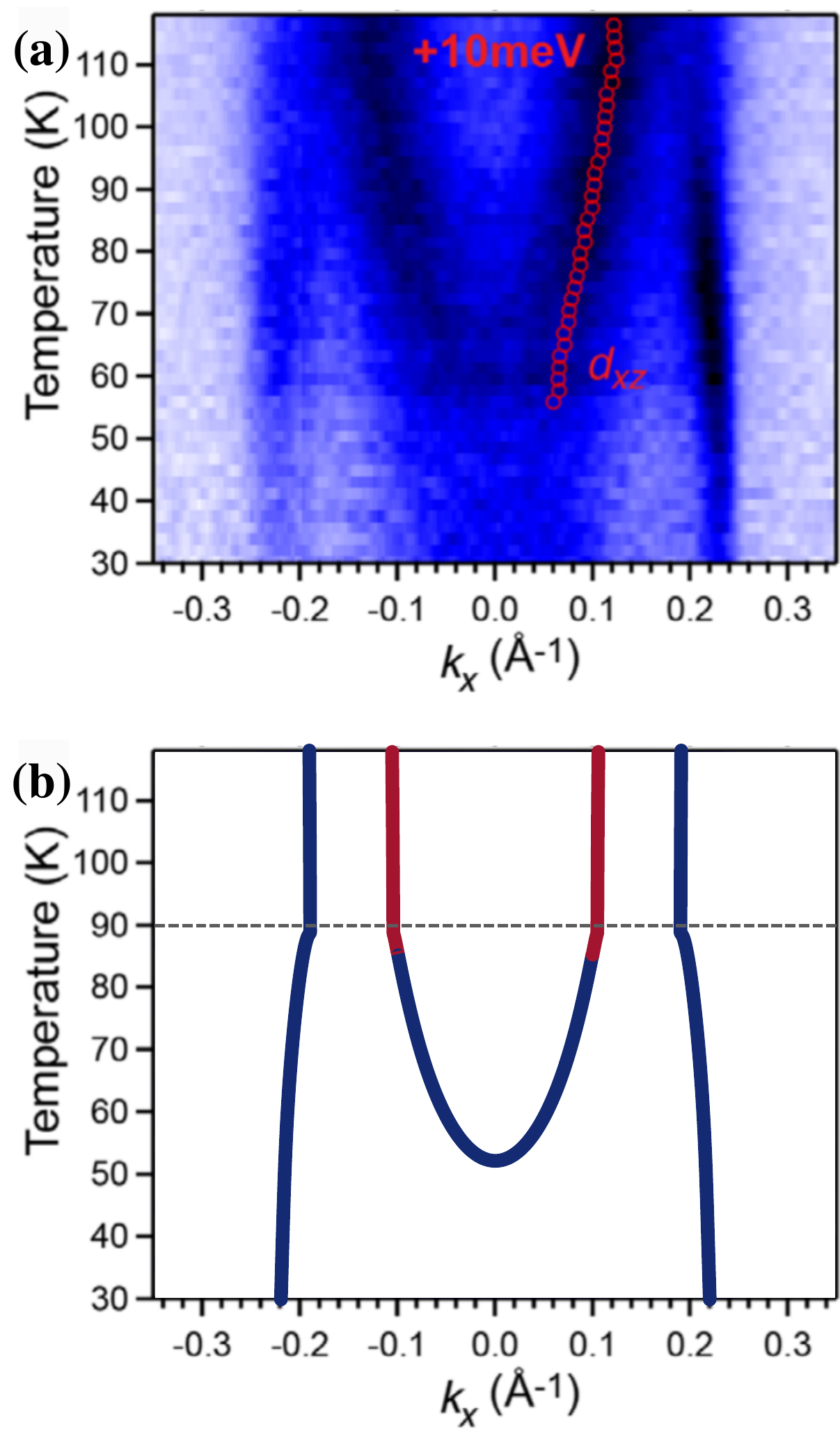}
			\caption{\textbf{Temperature dependent Liftshitz transition.} \textbf{a} ARPES measurements from Ref.~\cite{Yi2019}, Spectral intensity divided by the Fermi function taken at +10 meV above $E_{\text{F}}$ as a function of temperature. The disappearance of a band can be inferred at T=60~K. \textbf{b} Simulated evolution of the momentum of states near the $A$ point, along the $k_x$ axis $\kk=(k_x,\pi,\pi)$, which have an energy of $E=+10$~meV as a function of temperature.}
			\label{Fig:5}
		\end{figure}
        \begin{figure*}[t]
    	\centering
	    \includegraphics[width=1\linewidth]{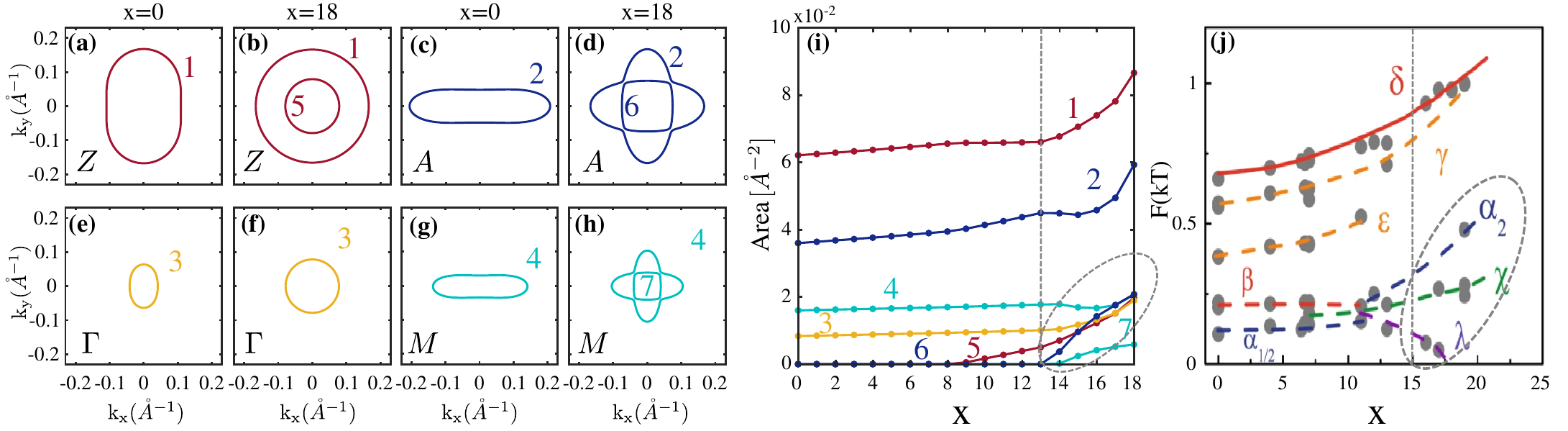}
    	\caption{\textbf{Simulation of Quantum Oscillation measurements.} \textbf{a-h} Fermi surfaces of FeSe$_{1-x}$S$_x$ in the orthohombic ($x=0$) and tetragonal ($x=18$ \%) phase. \textbf{i} Evolution of the Fermi surface areas as function of doping. Vertical dashed line marks the position of the Lifshitz transition at the zone corner within our parametrization. Note that we have ensured charge conservation by self consistently calculating the chemical potential. The areas of the hole pockets (labelled 1,3,5) and the electron pockets (labelled 2,4,6,7) preserve charge neutrality throughout. \textbf{j} Corresponding experimental data taken from Ref.~\cite{Coldea2019}. We have included a dashed circle in \textbf{i,j} to highlight the evidence for a re-entering $d_{xy}$ band. Note that we use a different color code in \textbf{a-i} than the authors of Ref.~\cite{Coldea2019} and that we  assign our pockets differently to the data and trend lines in \textbf{j}. }
	    \label{Fig:6}
        \end{figure*}
    Further evidence of a Lifshitz transition of the $d_{xy}$ orbital can be found in Ref.~\cite{Yi2019}. There, Yi. \textit{et. al.} reported temperature dependent ARPES measurements of the $A$ point of detwinned crystals of FeSe, where the orthorhombic domains were aligned along a given axis by uniaxial strain. When studying the temperature dependence of the spectral function around the $A$ point at +10~meV above the Fermi level, they report the observation of a shrinking inner electron band which disappears at this particular energy at approximately T = 60~K, as shown in \ref{Fig:5}(a). Our calculation of the momenta at which the electronic dispersion crosses the energy of +10~meV above the Fermi level is presented in Fig. \ref{Fig:5}(b). Indeed, it yields the same result as shown in Fig.~\ref{Fig:5}(a) with excellent agreement on temperature dependence and band position. This agreement between theory and experiment can be interpreted as a smoking gun for the $d_{xy}$ Lifshitz transition, however, we caution that there are some experimental technicalities, which can complicate the interpretation. For example, changing the temperature of a uniaxially strained system, as was performed in Ref. \cite{Yi2019}, may alter the relative ratio of the orthorhombic domain orientations, due to additional thermal contraction. This could imply that the disappearing intensity may instead arise from a change in population of orthorhombic domains, rather than a band shifting above $E_{\text{F}}$. Nevertheless, if confirmed by other experimental measurements, this could be the most direct evidence for a $d_{xy}$ Lifshitz transition. A similar observation has also been made for the sulphur doped FeSe$_{0.9}$S$_{0.1}$ system \cite{Cai2020,Cai2020b}.
	
	Alternate evidence for a Lifshitz transition at the zone corner, which does not require uniaxial strain, can be found by studying Quantum Oscillation measurements of the sulphur doped, FeSe$_{1-x}$S$_x$, systems. Isovalent sulphur doping is known to suppress the nematic state, which,  despite slightly changing the anion height of the lattice, leaves the underlying electronic structure effectively equivalent to that of the tetragonal state of FeSe \cite{Coldea2018,Reiss2017,Watson2015b}. Thus, by varying the sulfur content in FeSe$_{1-x}$S$_x$ it is possible to indirectly study the temperature dependence of the nematic order.  
	Recently, Shubnikov-de Haas oscillation measurements of the sulphur doped series have been reported by Coldea \textit{et. al.} \cite{Coldea2019}. These experiments detect extremal areas of Fermi surface pockets, which in FeSe corresponds to the 2D area of the hole and electron pockets at both $k_z = 0$ and $k_z = \pi$ respectively. 
	
    Previous interpretation of this experimental data was based on the assumption that two electron pockets are present in the nematic state. However, we find that our scenario, where there is only one electron pocket, is also qualitatively consistent with the quantum oscillation data if one assigns the $Z$ and $A$ pocket (labelled as 1 and 2 in Fig.~\ref{Fig:6}~(a,c)) to the $\delta$ and $\gamma$ frequencies of Ref. \cite{Coldea2019} and the $M$ and $\Gamma$ pockets (labelled as 3,4 in Fig.~\ref{Fig:6}~(e,g)) to the $\alpha$ and $\beta$ frequencies. Our scenario does not predict an additional $\epsilon$ frequency, which is however not detected in the experiment for the entire doping range.
    
    As a function of sulphur doping, a Lifshitz transition associated with the $d_{xy}$ band should lead to a sharp sudden change in extremal areas measured by this experimental technique. We illustrate this in Fig.~\ref{Fig:6}(i). By again assuming a mean field dependence of the nematic order parameter ($\Phi_i(x) = \Phi_i(0)\sqrt{1-(\frac{x}{x_0})}$, where $x_0$ = 18\%) we find that the four unique extremal areas present in the nematic state remain relatively consistent between 0\% and 13\% sulphur doping, with the inner hole pocket of the $Z$ point crossing the Fermi level at 10\%. However, at the Lifshitz transition of the $d_{xy}$ band, we find that all extremal areas sharply increase in magnitude and two additional extremal areas are generated. Bringing the total number of  unique extremal areas in the tetragonal state to seven. The reason for this sharp increase is twofold, the first is the sudden change of chemical potential as discussed in Fig.~\ref{Fig:4}(g), which mainly affects the hole pockets, and the second is the sudden change of the electron pockets from a two-fold symmetric, peanut-like, dispersion to a four-fold symmetric, petal-like, dispersion. Whilst the two largest areas corresponding to the $k_z=\pi$ hole pocket and the outer electron pocket are well separated from the $k_z=0$ pockets, the other five extremal areas are likely to have very similar (and very small) values. 
    
	Although a Lifshitz transition of the electron band was not the interpretation of Ref.~\cite{Coldea2019}, we observe that the data points measured for $x>15\%$ are in fact consistent with a re-entering $d_{xy}$ pocket at the $A$ and $M$ point, we display the experimental data of Ref.~\cite{Coldea2019} in Fig.~\ref{Fig:6}(j) for comparison. Very similar results can also be observed by suppressing nematic order under application of physical pressure to FeSe \cite{Reiss2020}.
	We note that the conversion of these simulated areas into units of Tesla using the Onsanger relation ($F = A\frac{\hbar}{2\pi e}$ where $A$ is the area) indicate that the simulated areas are approximately 70-80\% of that measured by experiment for all doping values, nevertheless the qualitative trends are consistent.

\subsection{Momentum Dependence of the superconducting gap}
    
    We now consider how the superconducting state will be affected by the presence of non-local $d_{xy}$ nematicity, and in particular how the gap structure will evolve from the nematic to tetragonal state.
	
	Superconductivity in the iron-based superconductors is believed to be driven by the interband repulsive interactions enhanced by spin fluctuations, and indeed the numerous comparisons between theory and experiment have shown that this assumption can correctly capture the symmetry of the superconducting gap \cite{Mazin2009,Hirschfeld2011,Chubukov2012}. The gap symmetry of the nematic state of FeSe has been a subject of intense experimental investigation, with multiple ARPES and scanning tunnelling microscopy (STM) measurements reporting a highly anisotropic momentum dependence of the superconducting gap \cite{Xu2016,Sprau2017,Rhodes2018,Kushnirenko2018,Liu2018,Hashimoto2018}. These measurements revealed regions in momentum space which are either nodal or posses a very small superconducting gap. Previously, it was shown that the solution of the linearised superconducting gap equation assuming a one-electron pocket Fermi surface and random phase approximation (RPA) correctly reproduced the experimental data in the nematic state \cite{Rhodes2018}, however this was achieved by phenomenologically removing the second electron pocket that was still present in the underlying band structure. In the scenario we have considered here, with sizeable $d_{xy}$ nematicity, the second electron pocket is shifted well above the Fermi level and we are now in a position to study how the momentum dependence of the superconducting gap evolves as nematicity is suppressed, and a $d_{xy}$ Lifshitz transition occurs,  which can be experimentally probed by measuring the momentum dependence of the superconducting gap as a function of sulphur doping.
	
	We utilize our orbital projected band model which allows for an analytical treatment of the pairing problem by solving mean-field BCS gap equations 
	self-consistently at zero temperature. Within the projected band model we adopt the interaction Hamiltonian of Ref.~\cite{Kang_s+id_2018} which combines Kohn-Luttinger and spin-fluctuation approaches. This leads to a projection on those pairing channels which contain intra-orbital $U_{\text{he}}$ and inter-orbital $J'_{\text{he}}$ and $J'_{\text{ee}}$ pairings that results in inter-band Cooper-pair hopping between hole and electron pockets, corresponding to $(\pi,0), (0,\pi)$ spin fluctuations,  and between both electron pockets, corresponding to $(\pi,\pi)$ fluctuations, respectively. 
	The interaction Hamiltonian then reads
	\begin{align}
	H_{\text{int}}&=U_{\text{he}}\sum_{\kk,\kk^\prime,\mu}d^\dagger_{\mu,\kk,\uparrow}d^\dagger_{\mu,-\kk,\downarrow}d_{\mu,-\kk^\prime+\Qv_\mu,\downarrow}d_{\mu,\kk^\prime+\Qv_\mu,\uparrow}\nn\\
	+&J'_{\text{he}}\sum_{\kk,\kk^\prime,\mu\neq\nu}d^\dagger_{\mu,\kk,\uparrow}d^\dagger_{\mu,-\kk,\downarrow}d_{\nu,-\kk^\prime+\Qv_\nu,\downarrow}d_{\nu,\kk^\prime+\Qv_\nu,\uparrow}\nn\\
	+&J'_{\text{ee}}\sum_{\kk,\kk^\prime,\mu\neq\nu}d^\dagger_{\mu,\kk+\Qv_\mu,\uparrow}d^\dagger_{\mu,-\kk+\Qv_\mu,\downarrow}d_{\nu,-\kk^\prime+\Qv_\nu,\downarrow}d_{\nu,\kk^\prime+\Qv_\nu,\uparrow}\nn\\
	&+h.c.\label{Eq:InteractionHamiltonian}
	\end{align}
    \begin{figure*}[t]
	\centering
	\includegraphics[width=1\linewidth]{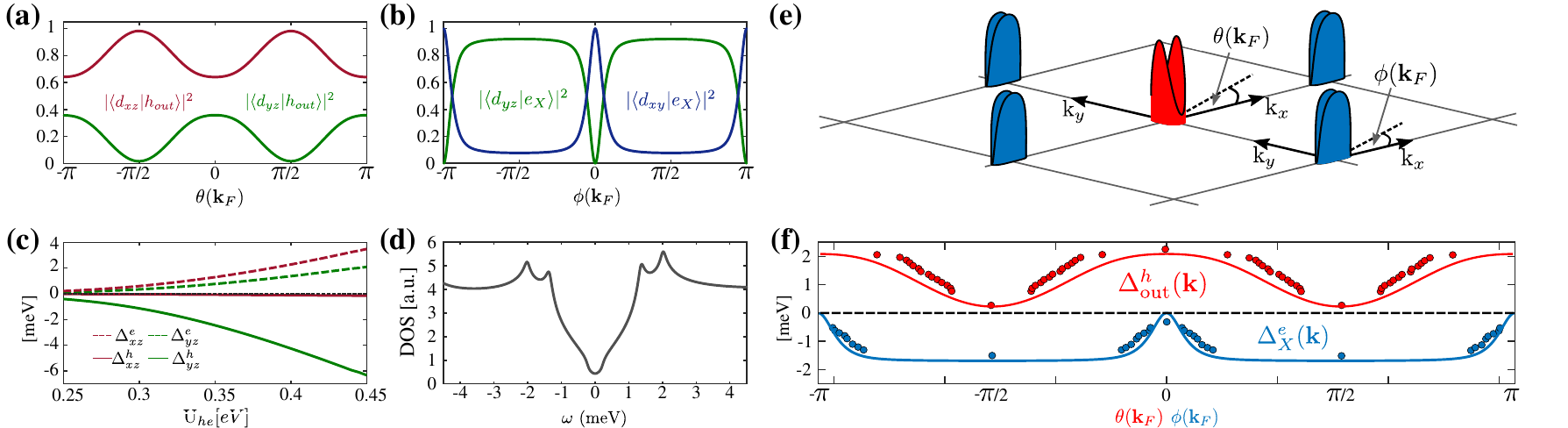}
	\caption{\textbf{Simulated momentum dependence of the superconducting gap of FeSe}. \textbf{a,b} Evolution of the orbital weight as a function of the Fermi angles $\theta(\kk_{\text{F}})$ and $\phi(\kk_{\text{F}})$ on the outer hole pocket and peanut shaped electron pocket respectively. \textbf{c} Solution of the self consistent gap equations, see Ref.~\cite{Supp}, as function of  $U_{\text{he}}$. \textbf{d} Superconducting density of states obtained for $U_{\text{he}}=431$~meV. \textbf{e} Magnitude of the superconducting gap in momentum space for $U_{\text{he}}=431$~meV, red and blue colours describe the positive and negative sign of the gap. Also shown are the definition of Fermi angles $\theta(\kk_{\text{F}})$ and $\phi(\kk_{\text{F}})$ at $Z$ and $A$ point, respectively, as measured with respect to $k_x$ direction. \textbf{f} Band gaps of the outer hole pocket (red) and peanut shaped electron pocket (blue) as function of the Fermi angle corresponding to \textbf{e}. The colored circles are data taken from quasiparticle interference experiments \cite{Sprau2017}. In all figures $J'_{\text{he}}=J'_{\text{ee}}=10$ meV. }
	\label{Fig:7}
    \end{figure*}	
    where $\mu,\nu\in\{xz,yz\}$. Even though our conclusions in the previous section suggest a sizeable contribution of $d_{xy}$ orbitals to nematicity, for superconducting pairing the $d_{xy}$ orbital is not so important. Superconductivity is dominated by inter-band, and intra-orbital, Cooper pair scattering. As there is no $d_{xy}$ weight on the hole pocket, the $d_{xy}$ orbital can only contribute to the superconducting pairing via scattering between two electron pockets ($J'_{\text{ee}}$), which is considerably weakened if only one electron pocket is present. In what follows we neglect the contribution of the $d_{xy}$ orbital to superconducting pairing. Later, in Sec. D, we will confirm this assumption by solving the full linerised gap equation for a 10-orbital tight binding model, and including the $d_{xy}$ orbital in the pairing. 
    

	We define superconductivity in orbital space, perform a mean-field decoupling and solve the BCS-gap equations together with the particle number equation self-consistently for the constant orbital pairing gaps assuming $U_{\text{he}}\gg J'_{\text{he}}=J'_{\text{ee}}$. In the gap equations we focus on the large hole pocket, the peanut shaped electron pocket and the incipient electron band above the Fermi level. We further neglect contributions from inter-band Cooper-pairing and integrate within an energy cutoff of 50~meV around the Fermi level. Further details are given in Ref.~\cite{Supp}. The superconducting gap on the outer hole pocket is given by
	\begin{align}
	\Delta^{h}_\text{\text{out}}(\kk)=\Delta^{h}_{xz}|a^{h}_{xz}(\kk)|^2+\Delta^{h}_{yz}|a^{h}_{yz}(\kk)|^2\label{Eq:HoleGap}  
	\end{align}
	with the constant orbital gaps $\Delta^{h}_{xz/yz}$ and the $\kk$-dependent orbital dressing factors $|a^{h}_{xz/yz}(\kk)|^2=|\langle d_{xz/yz}|h_\text{out}\rangle|^2$, which correspond to $d_{xz/yz}$ orbital content of the hole pockets, respectively. Assuming small SOC at the $A$ point the band-gap on peanut shaped pocket is given by in good approximation by
	\begin{align}
	\Delta^{e}_\text{out}(\kk)&\approx|a^{e}_{yz}(\kk)|^2\Delta^{e}_{yz}+|a^{e}_{xy}(\kk)|^2\Delta^{e}_{xy}\nn\\
	&\approx|a^{e}_{yz}(\kk)|^2\Delta^{e}_{yz}\label{Eq:PeanutGap}
	\end{align}
	, with $|a^{e}_{yz/xy}(\kk)|^2=|\langle d_{yz/xy}|e_X\rangle|^2$ the $d_{yz/xy}$ orbital weight on the peanut \cite{Supp}.
	
	In Fig.~\ref{Fig:7}(a) we show the orbital weight $|a^{h}_{xz/yz}(\kk)|^2$ on the hole pocket as function of the Fermi angle measured with respect to $k_x$ direction. In experiment, the gap at the hole pocket exhibits deep minima (and possibly accidental nodes) at $\theta(\kk_{\text{F}})=\pm\pi/2$ and has a maximum at $\theta(\kk_{\text{F}})=0$ resembling the $d_{yz}$ orbital weight. According to Eq.~(\ref{Eq:HoleGap}) this behaviour requires $\Delta^{h}_{xz}\ll\Delta^{h}_{yz}$. In Fig.~\ref{Fig:7}(b) we present the solution of the BCS gap equations as function of $U_{\text{he}}$, which yield the large ratio $\Delta^{h}_{yz}/\Delta^{h}_{xz}$ necessary to reproduce the correct angular dependence of $\Delta^{h}_\text{out}(\kk)$. A similar result for the hole pocket was also obtained in Ref.~\cite{Kang2018} where nematicity was solely given by $\Phi^{e}_{xz,yz}$. Within their model, however, two electron pockets were present at the Fermi level and they were not able to correctly reproduce the angular dependence of the gap at the peanut shaped electron pocket.  Ref.~\cite{Benfatto2018} also calculated Eq.~(\ref{Eq:HoleGap}) and Eq.~(\ref{Eq:PeanutGap}) for a model where nematicity was given solely by $\Phi^{e}_{xz,yz}$. However here they added an additional strong anisotropy to the superconducting pairing such that the second electron pocket did not contribute to superconductivity, but was still present at the Fermi level, in disagreement with ARPES measurements as discussed in Section B.
	
	The weight $|a^{e}_{yz/xy}(\kk)|^2$ of the $d_{xz}$ and $d_{xy}$ orbitals on the peanut shaped electron pocket is shown in Fig.~\ref{Fig:7}(b). Experimental results suggest that the angular dependence of the gap follows the $d_{yz}$ orbital weight of the peanut which has a minimum at $\phi(\kk_{\text{F}})=0$ and is almost isotropic elsewhere. This behaviour is captured by Eq.~(\ref{Eq:PeanutGap}).
	
	Furthermore, in Fig.~\ref{Fig:7}(c) we show that $|\Delta^{h}_{yz}|>|\Delta^{e}_{yz}|$, which leads to  $\text{max}(|\Delta^{h}_\text{out}(\kk_{\text{F}})|)>\text{max}(|\Delta^{e}_\text{out}(\kk_{\text{F}})|)$ as it is seen in experiment. In Fig.~\ref{Fig:7}(e) and (f) we show the order parameters $\Delta^{h}_\text{out}(\kk_{\text{F}})$ and $\Delta^{e}_\text{out}(\kk_{\text{F}})$ projected onto the Fermi surface and as a function of the Fermi angle and find excellent agreement with experimental data \cite{Xu2016,Sprau2017,Rhodes2018,Kushnirenko2018,Liu2018,Hashimoto2018}, of which we show data extracted from STM measurements of Ref.~\cite{Sprau2017}, depicted as red and blue circles in Fig~\ref{Fig:7}(e). For completeness, in Fig.~\ref{Fig:7}(d) we present the total density of states (DOS) $\rho(\omega)=-\frac{1}{\pi}\sum_{\kk}\text{Tr}\frac{\tau_0+\tau_3}{2}\text {Im}\hat{G}(\omega,\kk)$ calculated for the gap values that correspond to Fig.~\ref{Fig:7}(d). The DOS is nearly V-shaped, dedicated to the almost nodal gaps, and  exhibits two peaks at energies $\sim 1.5$ meV and $\sim 2.2$ meV, which correspond to the gap maximum of the electron and hole gap, respectively.
	
\begin{figure*}[t]
    	\centering
	\includegraphics[width=1\linewidth]{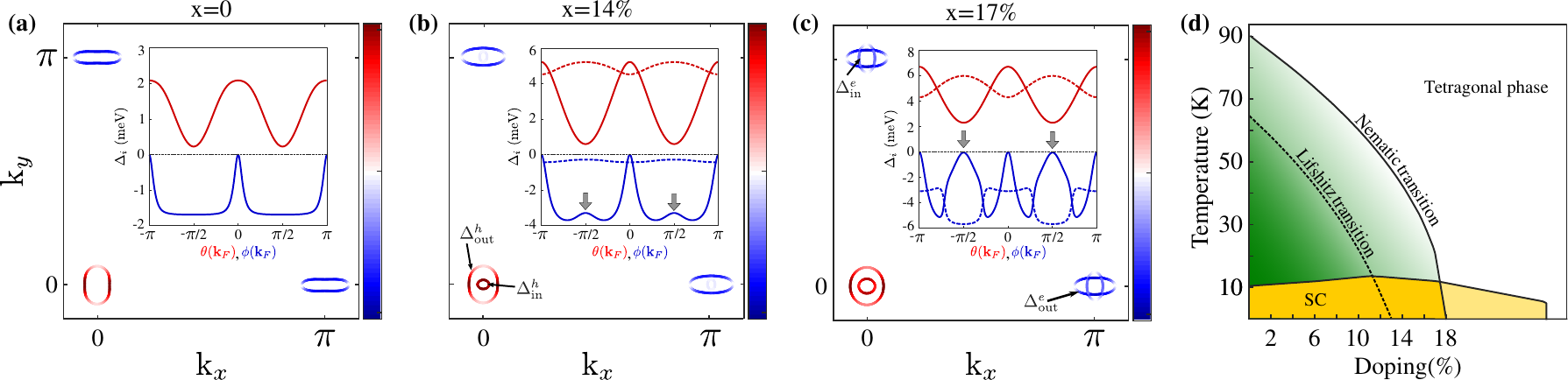}
	\caption{\textbf{Simulated effect of Sulphur doping on the momentum dependence of the superconducting gap}. \textbf{a-c} Momentum dependence of the gap for different values of Sulphur doping. The superconducting order parameter does not exhibit nodes and changes sign between electron and hole pockets. The arrows in \textbf{b-c} illustrates the main impact of the second electron pocket on the angular dependence of the gap on the outer and inner electron pockets near the $d_{xy}$ Lifshitz transition. Insets: Order parameters projected on the Fermi surface as function of the Fermi angle, measured with respect to $k_x$-direction at each high symmetry point. Results shown for outer and inner hole pocket in solid/dashed-red at $Z$ point and outer (peanut) and inner electron pocket in solid/dashed-blue at $A$ point. \textbf{d} Cartoon phase diagram of FeSe$_{1-x}$S$_x$ in the $d_{xy}$-nematic scenario as function of doping and temperature. The nematic phase is displayed in green whilst superconducting phase is yellow. The dashed line marks the Lifshitz transition of the inner electron pocket.}   
	\label{Fig:8}
    \end{figure*}

	Finally, in Fig.~\ref{Fig:8} we make a prediction about how the Fermi surface gap structure in the orthorhombic state of FeSe$_{1-x}$S$_x$ will evolve upon decreasing the nematic order parameter, for example as a function of increasing sulphur content. A schematic phase diagram highlighting the doping induced Lifshitz transition at the zone corner is shown in Fig.~\ref{Fig:8}(d). Here, we model the doping dependence of the nematic order parameters as $\Phi_i(x) = \Phi_i(0)\sqrt{1-(\frac{x}{x_0})}$, where $x_0$ = 18\%, and self consistently  calculate the pairing gaps as well as the chemical potential. Note, however, that at this stage we keep the interactions momentum-independent and do not alter its value as the doping increases and thus do not account for altering nesting conditions which will affect the pairing strength and gap size but not the momentum dependence. We observe that throughout the entire doping range the overall pairing symmetry exhibits a sign-change between the hole and electron pockets.
	
	At the same time, we find that within a scenario of the $d_{xy}$-dominant non-local nematicity there is a clear impact of the $d_{xy}$ Lifshitz transition on the angular dependence of the gap on the electron pockets as can been seen in Fig.~\ref{Fig:8}~(b,c). In particular, additional maxima/minima or accidental nodes for $x=0.17$ of the gap on the inner and outer electron pockets at $\theta = \pm \pi/2$ develop near the Lifshitz transition presumably due to enhanced ($\pi,\pi$) interaction between them. This could be an explanation for the two distinct superconducting gaps detected as a function of sulphur doping \cite{Hanaguri2018,Sato2018}. This specific angular dependence may be verified by ARPES/STM measurements and is one of the predictions of our $d_{xy}$-nematic scenario. Observe also that the angular dependence of the superconducting gaps in the tetragonal phase, $x>0.18$, is probably more complex due to recent observation of the time-reversal symmetry breaking \cite{Shibauchi2020}. 
    
	\subsection{Comparison with 10-orbital tight binding model}
	Throughout this manuscript we have used a minimal orbital projective model to describe the electronic dispersion around the Fermi level. To confirm the generality of our conclusions and to correctly account for $d_{xy}$ contribution to superconductivity we have also studied a 10-orbital tight binding model that has additionally been fit to ARPES data in the tetragonal state \cite{Rhodes2017}. We then use the same form of nematic order parameter which for the hole pockets can be written as
	
	\begin{equation}
	\begin{split}
	H^{\Gamma/Z}_{\text{nem}} = \Phi^{h}_{xz,yz}[d^\dagger_{yzA}d_{yzB}& -d^\dagger_{xzA}d_{xzB}](\cos(k_x)+\cos(k_y) \\&+ h.c.
	\end{split}
	\end{equation}
	and for the electron pockets can be written as
	
	\begin{align}
	H^{M/A}_{\text{nem}} = &\Phi^{e}_{xz,yz}[d^\dagger_{yz_A}d_{yz_B} +d^\dagger_{xz_A}d_{xz_B}](\cos(k_x)-\cos(k_y)) \nonumber \\
	&+d^\dagger_{xy_A}d_{xy_B}\Phi_4(\cos(k_x) - \cos(k_y) + h.c.
	\end{align}
	
	where $d^\dagger_{xy_{A/B}}$ ($d_{xy_{A/B}}$) is the creation (annihilation) operator for the $d_{xy}$ orbital on sublattice A or B in the two-atom unit cell respectively. Here we use the values $\Phi^{h}_{xz,yz}$=7~meV, $\Phi^{e}_{xz,yz}$ = -14~meV and keep $\Phi_4$ = 28~meV.
	
	To account for the Hartree term introduced in Eq.~\eqref{Eq:HartreeTerm} we additionally include the two equations
	\begin{equation}
	\begin{split}
	\Delta\epsilon_{xy}&[d^\dagger_{xy_A}d_{xy_B}(\cos(k_x) + \cos(k_y) \\ -&d^\dagger_{xy_A}d_{xy_A}(\cos(k_1) + \cos(k_2)] + h.c.,
	\end{split}
	\label{Eq:10Orbit_Hartree}
	\end{equation}
	This term ensures the asymmetric splitting of the two $d_{xy}$ bands around the $M$ ($A$) point are captured whilst leaving the additional $d_{xy}$ band, located around -50~meV at the $\Gamma$ ($Z$) point, unaffected. As with the case of the orbital projective model, Eq.~\eqref{Eq:10Orbit_Hartree} does not break any additional symmetries, or change the Fermi surface topology, however it is required to reproduce the fine details of the band dispersions below the Fermi level. We set $\Delta\epsilon_{xy} = 28$~meV within this formalism. We present the Fermi surface and band dispersion for this model in Ref.~\cite{Supp} showing nearly identical band dispersions to the orbital projective model of Fig.~\ref{Fig:1}.
	
	We now study the momentum dependence of the superconducting gap predicted from this model. To do this we solve the standard linearized superconducting gap equation \cite{Graser2009},
	\begin{equation}
	\label{eigenvalue equation}
	-\frac{1}{4\pi^2}\sum_{\nu}\oint_{C_\nu}\frac{d\mathbf{k'}_\parallel}{v_F(\mathbf{k'})}\Gamma_{\mu\nu}(\kk,\mathbf{k'}) g_\alpha(\mathbf{k'}) = \lambda_\alpha g_\alpha(\kk).
	\end{equation}
	
	Here, we integrate the pairing vertex, $\Gamma_{\mu\nu}(\kk,\mathbf{k'})$ over all $\mathbf{\mathbf{k'}}$ states on the Fermi surface for each band, $\nu$. The eigenvectors corresponding to the largest eigenvalue then describe the symmetry of the leading superconducting instability suggested by the pairing vertex. Importantly, within this framework we treat all orbitals, including the $d_{xy}$ orbital, as equally coherent \footnote{Although in principle the quasiparticle weight of each orbital can be different, here we assume that the difference is small enough that it can be considered negligible}.
	
	We use a pairing vertex that describes spin-fluctuation mediated superconductivity utilising the random phase approximation (RPA) in the presence of spin orbit coupling \cite{Saito2015,Rhodes2018}. The exact form of the pairing vertex is rather long and left to the appendix, however the important properties that control this pairing vertex are i) the orbital content present at the Fermi level and ii) the form of the spin susceptibility used to generate pairing. 
	
	One has to bear in mind, however, that the RPA approximation is only valid in the regime where $\frac{\omega_{\text{sf}}}{E_{\text{F}}} < 1$, where $\omega_{\text{sf}}$ is the spin fluctuation frequency and $E_{\text{F}}$ is the Fermi energy \cite{Yamakawa2017}, which is not true for FeSe where the Fermi energy is $<$10~meV \cite{Watson2015}.
	
	However, the exact details of the spin susceptibility, and pairing vertex, are less important to the superconducting properties of FeSe than the available states present around the Fermi level. To emphasise this, we solve the linearised superconducting gap equation using two distinct forms of the spin susceptibility. In the first we employ the RPA pairing vertex, calculated explicitly from the itinerant electronic structure in the tetragonal and nematic state,
	\begin{align}
	\chi^0_{pq;st}(\mathbf{q},\omega) = -\frac{1}{N}\sum_{\kk,\mu\nu}  \frac{a^s_\mu(\kk)a^{p*}_\mu(\kk)a^q_\nu(\kk+\mathbf{q})a^{t*}_\nu(\mathbf{k+q})}{\omega + E_\nu(\mathbf{k+q}) - E_\mu(\kk)}\nn\\
	\times[f(E_\nu(\mathbf{k+q})) -f(E_\mu(\kk))].
	\label{Eq:Lindhard equation}
	\end{align}
	where, $a^s_\mu(\kk)$ describe the eigenvector of orbital $s$ and band $\mu$ at momentum $\kk$, $E_\mu(\kk)$ is the corresponding eigenvalue and $f(E)$ is the Fermi function calculated at T=10~K in the nematic state and 100~K in the tetragonal state.
	
	In the second form, we employ a greatly simplified phenomenological form of the spin susceptibility, 
	\begin{equation}
	\chi^0(\mathbf{q},\omega) = \sum_{i} \frac{X_{\mathbf{Q}_i}}{1+\xi^2(\mathbf{q}-\mathbf{Q}_i)^2 - i(\omega/\Omega_{\text{max}})}, 
	\label{Eq:SpinSuscep}
	\end{equation}
	such as has been used to study the cuprates and other iron-based superconductors \cite{Mills1990,Heimes2012}. Here, $\mathbf{Q}_i$ are the antiferromagnetic wavevectors $(\pi,0)$ and $(0,\pi)$ and $X_{\mathbf{Q}_i}$ is a constant proportional to the real part of the spin response at the corresponding {\bf Q}$_i$, with units $eV^{-1}$, which we set to 1 for each  {\bf Q}$_i$ for illustrative purposes. 
	
\begin{figure}[t]
	\centering
	\includegraphics[width=1\linewidth]{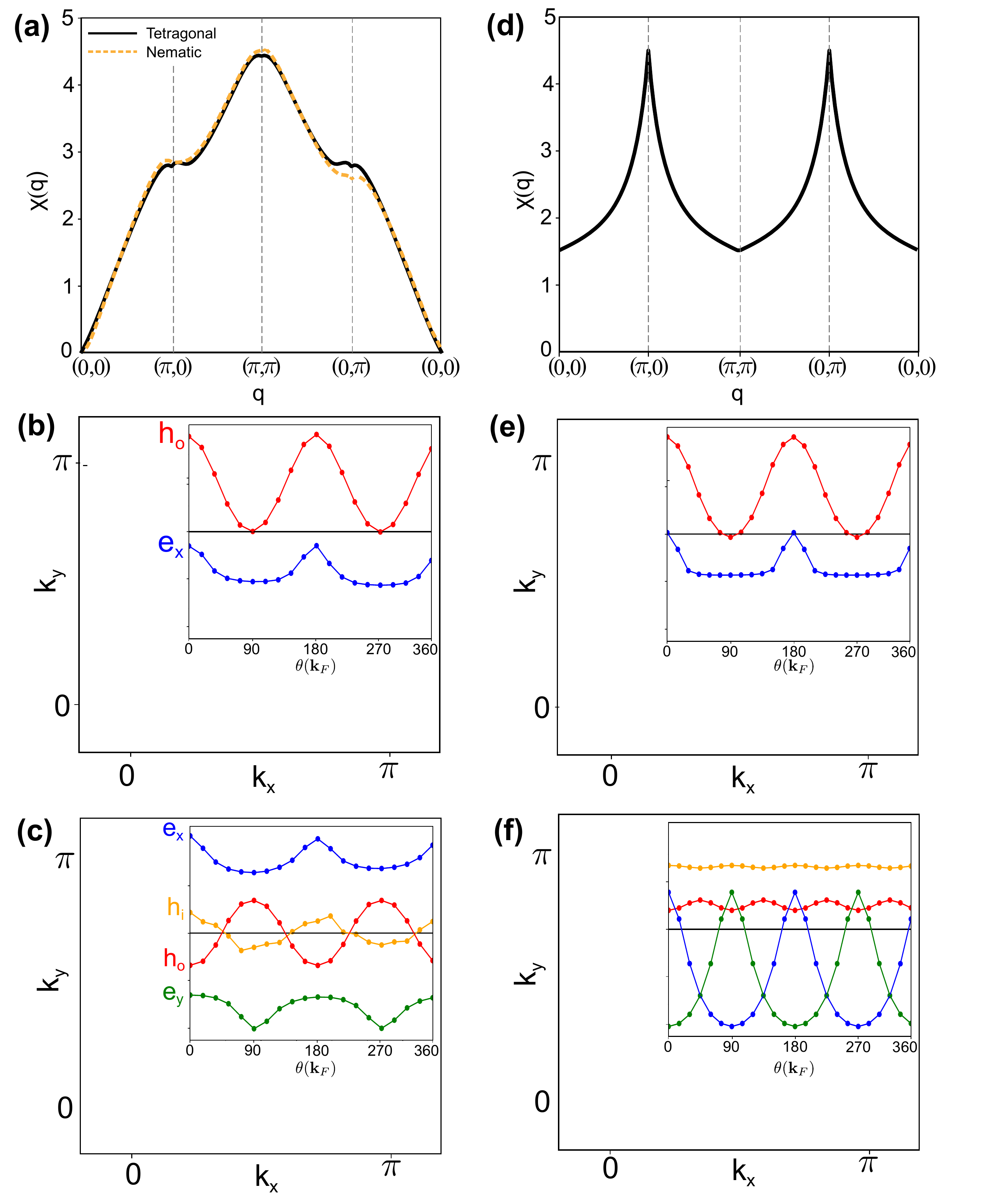}
	\caption{\textbf{Comparison of the superconducting gap simulations with a 10-orbital tight binding model}. \textbf{a} Spin susceptibility calculated using the itinerant spin susceptibility defined in Eq.~\eqref{Eq:Lindhard equation}. \textbf{b,c} Eigenvectors of the leading eigenvalue from Eq.~\eqref{eigenvalue equation} calculated using the itinerant spin susceptibility in \textbf{b} the nematic state and \textbf{c} The tetragonal state. The inset defines the angular dependence of the gap for each pocket in arbitrary units, taken anticlockwise from the $k_x$-axis. $e_x$ and $e_y$ define the momentum dependence of the gap at $(\pi,0)$ and $(0,\pi)$ respectively, whilst $h_o$ and $h_i$ describe the inner and outer hole pocket. \textbf{d-f} Equivalent results obtained from the phenomenological spin susceptibility defined in Eq.~\eqref{Eq:SpinSuscep}. Here we set U=0.5~eV, $U'=\frac{U}{6}$,$J=0.1U$ and $J'=J$ throughout and use the unfolded 1-Fe unit cell \cite{Rhodes2017}. Further details can be found in the supplemental material.}
	\label{Fig:9}
    \end{figure}

	We additionally set $\xi$ (a measure of the correlation length of the spin fluctuations) to unity and neglect inter-orbital contributions in this approximation. The spin susceptibility calculated using Eq.~\eqref{Eq:Lindhard equation} and Eq.~\eqref{Eq:SpinSuscep} are presented in Fig.~\ref{Fig:9}(a) and (d) respectively. We observe that the itinerant form of the spin susceptibility predicts maximum spin fluctuations to occur at $(\pi,\pi)$, in agreement with previous models of the tetragonal state \cite{Mukherjee2015,Kreisel2018,Kreisel2015,Kreisel2017}, and that the nematic order parameter only induces a slight increase (decrease) to the spin susceptibility at $(\pi,0)$ ($(0,\pi)$). Although this form of spin susceptibility does not correctly describe the fully dynamics of the spin response as seen in inelastic neutron scattering experiments \cite{Wang2015,Chen2019} we find that remarkably, the correct highly anisotropic momentum dependence of the superconducting gap is still obtained as the leading solution to the linearised gap equation of the nematic state, as shown in Fig.~\ref{Fig:9}(b). This is equivalently the case for the phenomenological spin susceptibility in Fig.~\ref{Fig:9}(e). The reason for this agreement is that,  due to the Fermi surface topology, superconducting pairing in the nematic state is controlled by approximately only two scattering vectors $\Qv \approx (0,0)$ and $\Qv \approx (\pi,0)$, such that the rest of the spin susceptibility structure does not impact the superconducting properties.
	
	However, in the nominally tetragonal state, with two electron pockets, we find that the form of spin susceptibility does have an effect on the leading pairing symmetry. The itinerant pairing vertex, with maximum spin fluctuations at $(\pi,\pi)$, suggests $d$-wave superconductivity will be the leading instability, as shown in Fig.~\ref{Fig:9}(c). However, if $(0,\pi)$ and $(\pi,0)$ fluctuations are assumed to be dominant then $s^{\pm}$ becomes the leading instability (Fig.~\ref{Fig:9}(f)).
	
	Additionally, we find that the only contribution of the $d_{xy}$ orbital in superconducting pairing is the possible generation of accidental nodes at the tips of the electron pockets, which verifies that the assumption made in Sec. C is valid. This observation  emphasizes the robustness of our prediction for the impact of the Lifshitz transition of one electron band due to $d_{xy}-$nematicity on the superconducting gap near the nematic to tetragonal transition. 

\section{Discussion} 
	The $d_{xy}$ orbital has been largely underestimated within theories of the nematic state for the iron-based superconductors. With most theories focusing solely on the degeneracy lifting of the $d_{xz}$ and $d_{yz}$ orbitals. Although, the possibility of anisotropic hopping between the $d_{xy}$ orbitals of nearest neighbour atoms have been also considered \cite{Xing2017,Vafek2018,Christensen2020}, it was thought to have a minor role for the electronic structure modification in the nematic state. The resulting contradiction of theoretically predicted Fermi surface topology in the nematic state with that experimentally measured by ARPES had previously been attributed to the incoherence of the $d_{xy}$-dominated electron pocket. This was indirectly supported by DMFT calculations, arguing about the stronger correlations effect on the $d_{xy}$ orbital \cite{Yin2011}, yet we are not aware of DMFT calculations that would correctly reproduce the measured one-electron pocket Fermi surfaces of the nematic state of FeSe. In order to generate the experimentally observed Fermi surface, we adopted another scenario where the sizeable $d_{xy}$ nematic order parameter is larger than the $d_{xz}/d_{yz}$ term and find that it allows consistent explanation of several existing experimental controversies. In addition, this scenario predicts a Lifshitz transition of one electron band in the sulphur doped FeSe$_{1-x}$S$_x$ systems, which should occur prior to the orthorhombic to tetragonal transition. This may explain recent resistivity measurements in high magnetic field which have reported quantum-critical behaviour at a sulphur doping value of $x=0.16$ \cite{Licciardello2019}, two percentage points lower than the disappearance of the nematic state which occurs at $x=0.18$ \cite{Bohmer2017}, as well as the sudden change in superconducting gap observed in specific heat and STM measurements \cite{Sato2018,Hanaguri2018}. Assuming that the nominal doping values are correct between different experiments, we expect that the observed quantum critical behaviour is associated with the Lifshitz transition predicted here.
	
	Apart from the interest on its own, the sizeable $d_{xy}$-nematicity poses further constraints for microscopic theories of nematicity. The anisotropic hopping pathway of the $d_{xy}$ orbital, coupled with the experimental observation of a momentum dependent $d_{xz}/d_{yz}$ rearrangement suggest that nematicity is a strongly non-local phenomena (likely of magnetic rather than purely orbital origin) and may be governed by nearest neighbour interactions \cite{Schattner2016}. Observe also that this order needs to be re-examined within the RG weak-coupling analysis performed previously\cite{Claessen2017,Xing2017,Xing2018}. It is also interesting to note that local ab-initio calculations, such as LDA, LDA + DMFT, and even QSGW+DMFT are unable to account for such a prominent evolution of the bands upon entering the nematic state \cite{Fedorov2016,Watson2017c,Acharya2020_arXiv}. Although a recent DFT calculation proposed an E$_{\text{u}}$ symmetry nematic order parameter, using symmetry preconditioned wavefunctions \cite{Long2020}, that generates an electronic structure where one of the electron pockets is above the Fermi level. The lack of a consistent ab-initio description of the nematic state, coupled with our observation of a maximum nematic order parameter on the most correlated $d_{xy}$ band strongly suggests that nematicity arises from a non-local strongly-correlated phenomenon. It is also interesting to note that Te doping induces an orbital selective Mott transition of the $d_{xy}$ orbital, and suppresses the nematicity  \cite{Jiang2020_arXiv,Huang2020_arXiv}.
	
	The scenario of the $d_{xy}$ dominant nematicity has also important consequences for alternative theoretical explanations for experimental properties of FeSe, such as the need for highly anisotropic orbital-selective quasiparticle weights or spin fluctuations to describe the momentum dependence of the superconducting gap \cite{Kreisel2017,Sprau2017,Benfatto2018,Hu2018,Cercellier2019}. These theories were born from the observation that a two-electron pocket Fermi surface could not correctly describe experimental measurements, and that some form of anisotropy was required, either within the quasiparticle weight of the $d_{xz}$ and $d_{yz}$ orbitals, or in the direction of the spin fluctuations. Importantly, both ideas have the consequence that they suppress any contribution of a second electron pocket from superconducting pairing. What we have shown here is that the discrepancies between theory and experiment can be simply explained as a consequence of starting with an incorrect Fermi surface topology.  This point has been also argued in several publications \cite{Rhodes2018,Rhodes2019}, however, in these papers a justification for the removal of the second electron pocket was lacking. Here, we have presented a scenario which can justify the removal of the second electron pocket, by assuming large nematic ordering in the $d_{xy}$ orbital.
	
	 We want to point out that the addition of a Hartree shift alongside the nematic order parameters is necessary in our model to reproduce the asymmetric nematic splitting of the $d_{xy}$ band, suggested by experiments \cite{Yi2019}, which constrains the position of the $d_{xy}$ dominated bands far below the Fermi energy. Future work is required to elucidate the potential link of this term with nematicity.
	
	To conclude, we have studied a model of FeSe and FeSe$_{1-x}$S$_x$ which exhibits dominant nematic order in the $d_{xy}$ channel and have shown that this assumption can naturally explain a multitude of  unusual experimental properties observed within this system. This finding greatly constrains the possible mechanisms that can be attributed to the phenomena of nematicity. 

Note added: During the submission process of our manuscript we became aware of a theoretical paper by Steffensen \textit{et. al.} \cite{Steffensen2020_arXiv} also addressing the question of a single electron pocket Fermi surface of FeSe.

\textbf{Acknowledgements} We thank Andrey Chubukov, Yuji Matsuda, Taka Shibauchi, Peter Wahl, Amalia Coldea, Steffen Boetzel, Matthew Watson and Timur Kim for insightful conversations. LCR acknowledges funding from the Royal Commission for the Exhibition of 1851. JB, MAM, and IME. gratefully acknowledge financial support by the German Research Foundation within the DFG Project ER463/14-1. 

\textbf{Competing Interests} The Authors declare no Competing Financial or Non-Financial Interests.

\textbf{Data Availability} The authors declare that all data supporting the findings of this study are available within the paper and its supplementary information files.

\textbf{Author Contribution} LCR and JB conceptualised the project. JB, MAM and IE performed calculations using the orbital projective model. LCR and ME performed calculations using the 10-orbital tight binding model. All authors contributed to writing the manuscript. 

%

\appendix
\begin{widetext}
\section*{Supplementary note 1}
In this note, we discuss \textbf{the parametrization of  states near} $\mathbf{\Gamma}$ \textbf{and} $\mathbf{Z}$ \textbf{point}.
We model the fermionic states at the center of the Brillouin zone using the effective low energy  model introduced in Ref.~\cite{Cvetkovic13}. In the tetragonal as well as in the orthorhombic phase the states near the Fermi level at the $\Gamma$ and $Z$ point can be described in terms of a two component spinor $\Psi^\dagger_{\Gamma/Z,\sigma}(\kk)=(d^\dagger_{xz,\sigma}(\kk), d^\dagger_{yz\sigma}(\kk))$ with momentum $\kk$ measured with respect to $(0,0,0)$ at $\Gamma$ and $(0,0,\pi)$ at $Z$ point,respectively The nematic order parameter, which we discuss in the main text, and spin-orbit coupling enter the parametrization via
\begin{align}
H_\text{nem}^{\Gamma/Z}=&\sum_{\kk\sigma}\Phi^{h}_{xz,yz}\Psi^\dagger_{\Gamma/Z,\sigma}(\kk)\tau_3\Psi_{\Gamma/Z,\sigma}(\kk)\\
H^{\Gamma/Z}_{\text{SOC}}&=\sum_{\kk\sigma,\sigma'}\frac{\lambda^{h}_{\text{SOC}}}{2}\Psi^\dagger_{\Gamma/Z,\sigma}(\kk)\tau_2\Psi_{\Gamma/Z,\sigma^{\prime}}(\kk)\sigma_3^{\sigma\sigma'}
\end{align}
Written in coordinates of the \textit{unfolded} Brilloin zone the Hamiltonian is given by
\begin{align}
H_{\Gamma/Z,\sigma}(\kk)&=\left[\epsilon^{h}_{xz,yz}-\mu-\frac{k^2}{2m_{h}}\right]\tau_0
-\left[\frac{b}{2}(k_x^2-k_y^2)-\Phi^{h}_{xz,yz}\right]\tau_3
-2ck_xk_y\tau_1
+\sigma\frac{\lambda^{h}_{\text{SOC}}}{2}\tau_2
\nn\\
&=\left[\epsilon^{h}_{xz,yz}-\mu-\frac{k^2}{2m_{h}}\right]\tau_0-\left[\frac{b}{2}k^2\cos(2\theta)-\Phi^{h}_{xz,yz}\right]\tau_3+\frac{b}{2}k^2\sin(2\theta)\tau_1 +\sigma\frac{\lambda^{h}_{\text{SOC}}}{2}\tau_2,\label{Eq:Hamiltonian Gamma/Z}
\end{align}
where we used $c=-\frac{b}{2}$ and $\mu$ is the chemical potential that fixes the number of particles accounting for Luttinger theorem. Exact diagonalization leads the band dispersions
\begin{align}
\xi^{h}_\text{out/in}(\kk)=\epsilon^{h}_{xz,yz}-\mu-\frac{k^2}{2m_{h}}\pm\sqrt{(\frac{\lambda^{h}_{\text{SOC}}}{2})^2+(\Phi^{h}_{xz,yz})^2+\frac{b^2}{4}k^4-2\Phi^{h}_{xz,yz}\frac{b}{2}k^2\cos(2\theta)}
\end{align}
corresponding to the larger outer and smaller inner hole pocket $h_\text{in}$ and $h_\text{out}$, respectively. Note that the inner hole pocket sinks below the Fermi level once $\epsilon^{h}_{xz,yz}-\mu<\sqrt{(\frac{\lambda^{h}_{\text{SOC}}}{2})^2+(\Phi^{h}_{xz,yz})^2}$.\newline\newline
Within our model we define superconductivity as momentum independent mean field order parameters in orbital space. The order parameters for $d_{yz}$ and $d_{xz}$  intra-orbital onsite pairing are given by
\begin{align}
\Psi^\dagger_{\Gamma/Z,\uparrow}(\kk)\left(
\begin{array}{cc}
\Delta^{h}_{xz} &  \\
& \Delta^{h}_{yz}
\end{array}
\right)\Psi^*_{\Gamma/Z,\downarrow}(-\kk)
\end{align}	

which in Nambu notation $\Psi^\dagger_\kk=(\Psi^\dagger_{\Gamma/Z,\uparrow}(\kk), \Psi^T_{\Gamma/Z,\downarrow}(-\kk) )$ yields the BdG Hamiltonian $H^{\Gamma/Z}_{\text{\text{\text{BdG}}}}=\sum_\kk\Psi^\dagger_\kk H^{\Gamma/Z}_{\text{\text{\text{BdG}}}}(\kk)\Psi_\kk$ in orbital space
\begin{align}
H^{\Gamma/Z}_{\text{\text{\text{BdG}}}}(\kk)=\left(\begin{array}{cc}
H_{\Gamma/Z,\uparrow}(\kk)&\hat{\Delta}_{h}  \\
\hat{\Delta}_{h}^\dagger&-H_{\Gamma/Z,\uparrow}(\kk) 
\end{array}
\right)\label{Eq:HBdG Gamma Orb Basis}
\end{align}
with the pairing matrix $\hat{\Delta}_{h}=\frac{\tau_0+\tau_3}{2}\Delta^{h}_{xz}+\frac{\tau_0-\tau_3}{2}\Delta^{h}_{yz}$. We perform a unitary transformation 
\begin{align}
   \hat{U}^\dagger_\kk =\left(\begin{array}{cc}
	\alpha^{h}_{xz}(\kk) & \alpha^{h}_{yz}(\kk) \\ 
	-\alpha^{h*}_{yz}(\kk)& \alpha^{h*}_{xz}(\kk)
\end{array} 
\right)=
\left(\begin{array}{cc}
	\sin(\phi^h_\kk)e^{i\alpha} & \cos(\phi^h_\kk) \\ 
	-\cos(\phi^h_\kk) & \sin(\phi^h_\kk)e^{-i\alpha}
\end{array} 
\right)
\end{align}
that diagonalizes $H_{\Gamma/Z,\uparrow}(\kk)$ from orbital into band space and obtain the BdG-Hamiltonian band basis
\begin{align}
\hat{U}^\dagger_\kk H^Z_{\text{\text{\text{BdG}}}}\hat{U}_\kk=\left(\begin{array}{cccc}
\xi^{h}_{\text{in}}(\kk) &  & \Delta^{h}_{\text{in}}(\kk) & \Delta^{h}_{h_1h_2}(\kk) \\
& \xi^{h}_{\text{out}}(\kk) & \Delta^{h}_{h_2h_1}(\kk) & \Delta^{h}_{\text{out}}(\kk) \\
\Delta^{h*}_{\text{in}}(\kk) & \Delta^{h*}_{h_2h_1}(\kk) & -\xi^{h}_{\text{in}}(\kk) &  \\
\Delta^{h*}_{h_1h_2}(\kk) &\Delta^{h*}_{\text{out}}(\kk) &  & -\xi^{h}_{\text{out}}(\kk)
\end{array}
\right)
\end{align}
in which the intra- and inter-band gaps are given by
\begin{align}
\Delta^{h}_\text{\text{out}}(\kk)&=\Delta^{h}_{xz}\sin^2(\phi^h_\kk)+\Delta^{h}_{yz}\cos^2(\phi^h_\kk),\quad
\Delta^{h}_\text{\text{in}}(\kk)=\Delta^{h}_{xz}\cos^2(\phi^h_\kk)+\Delta^{h}_{yz}\sin^2(\phi^h_\kk),\\
\Delta^{h}_{h_1h_2}(\kk)&=\frac{1}{2}\sin(2\phi^h_\kk)e^{i\alpha}(\Delta^{h}_{yz}-\Delta^{h}_{xz}),\quad
\Delta^{h}_{h2h_1}(\kk)=\frac{1}{2}\sin(2\phi^h_\kk)e^{-i\alpha}(\Delta^{h}_{yz}-\Delta^{h}_{xz}),\label{Eq:GammaInterGap}
\end{align}
respectively, and
\begin{align}
|\alpha^h_{xz}(\kk)|^2=\sin^2(\phi^h_\kk)=\frac{1}{2}\left[1+\frac{\Phi^{h}_{xz,yz}-\frac{b}{2}k^2\cos(2\theta)}{\sqrt{(\frac{\lambda^{h}_{\text{SOC}}}{2})^2+(\Phi^{h}_{xz,yz})^2+\frac{b^2}{4}k^4-2\Phi^{h}_{xz,yz}\frac{b}{2}k^2\cos(2\theta)}}\right],\\
|\alpha^h_{yz}(\kk)|^2=\cos^2(\phi^h_\kk)=\frac{1}{2}\left[1-\frac{\Phi^{h}_{xz,yz}-\frac{b}{2}k^2\cos(2\theta)}{\sqrt{(\frac{\lambda^{h}_{\text{SOC}}}{2})^2+(\Phi^{h}_{xz,yz})^2+\frac{b^2}{4}k^4-2\Phi^{h}_{xz,yz}\frac{b}{2}k^2\cos(2\theta)}}\right],
\end{align}
denotes the orbital content. 
We fit the dispersion to ARPES data at the $\Gamma$ and $Z$ point at 10 K in the orthohombic state. The fitting parameters are listed in Tab.~\ref{Tab.FittingZ}.

\begin{table}[h]
	\begin{tabular}{c|c c c c}	
		& $\Gamma$& $Z$ &  \\ 
		\hline 
		$\epsilon^{h}_{xz,yz}$ &-8&$ 12 $  & &$\text{meV}$ \\ 
		$\frac{1}{2m_{h}}$ & $4730$&$ 2841 $  & &$\text{meV}\enspace\mathring{A}^2$ \\ 
		
		$b$ &$4664$&$2801.3 $ & &$\text{meV}\enspace\mathring{A}^2$ \\ 
		
		$c$ &$-2332$&$-1400.7 $ & &$\text{meV}\enspace\mathring{A}^2$ \\ 
		
		$\Phi^{h}_{xz,yz}$ &$15$&$ 15 $ & &$\text{meV}$ \\ 		
		
		$\lambda^{h}_{\text{SOC}}$ &$23$&$23$ & &$\text{meV}$ \\ 
	\end{tabular} 		 
	\caption{Fitting parameters for the states at the $Z$ point fitted to ARPES data at 10 K.\label{Tab.FittingZ}}
\end{table}

\section*{Supplementary note 2}
In this note, we discuss \textbf{the parametrization of  states near the} $\mathbf{M}$ \textbf{and} $\mathbf{A}$ \textbf{point}, which can be described in terms of the four component spinor $\Psi^\dagger_{M/A,\sigma}(\kk)=\left(\Psi^\dagger_{X,\sigma}(\kk),\Psi^\dagger_{Y,\sigma}(\kk)\right)$, where $\Psi^\dagger_{X,\sigma}(\kk)=(d^\dagger_{yz,\sigma}(\kk+\Qv_X), d^\dagger_{xy,\sigma}(\kk+\Qv_X))$ and $\Psi^\dagger_{Y,\sigma}(\kk)=(d^\dagger_{xz,\sigma}(\kk+\Qv_Y), d^\dagger_{xy,\sigma}(\kk+\Qv_Y))$. At the $M/A$ point, in the folded Brillouin zone, $X$ and $Y$ pockets are folded upon each other and from now on we understand $\kk$ as the deviation from $(\pi,\pi,0)$ and $(\pi,\pi,\pi)$ at $M$ and $A$ point, respectively. The nematic orderparameter, the Hartree term (both discussed in the main text) and spin-orbit coupling enter the parametrization via
\begin{align}
H_{\text{nem}}^{A/M}&= \sum_{\kk} 
\Psi^\dagger_{M/A,\sigma}(\kk)\left[-\Phi^{e}_{xz,yz}\left(\tau_3\otimes\frac{\tau_0+\tau_3}{2}\right)-\Phi_{4}\left(\tau_3\otimes\frac{\tau_0-\tau_3}{2}\right) \right]\Psi_{M/A,\sigma}(\kk)
\\
H^0_{\text{Hartree}}&=\sum_{\kk} 
\Psi^\dagger_{M/A,\sigma}(\kk)\left[\Delta\epsilon_{xy}\left(\tau_0\otimes\frac{\tau_0-\tau_3}{2}\right)  \right]\Psi_{M/A,\sigma}(\kk)\\
H^{M/A}_\text{SOC}&=\sum_{\kk}\Psi^\dagger_{M/A,\sigma}(\kk)\left(\begin{array}{cc}
0 & h^{\sigma,\sigma^\prime}_{\text{SOC}} \\ 
\left(h^{\sigma,\sigma^\prime}_{\text{SOC}}\right)^\dagger & 0
\end{array} \right)\Psi_{M/A,\sigma}(\kk)\\
h^{\sigma,\sigma^\prime}_{\text{SOC}}&=i\frac{\lambda^{e}_{\text{SOC}}}{2}\left(\frac{\tau_1+i\tau_2}{2}\otimes\sigma^x_{\sigma,\sigma^\prime}+\frac{\tau_1-i\tau_2}{2}\otimes\sigma^y_{\sigma,\sigma^\prime}\right)
\end{align}
Written in coordinates of the \textit{unfolded} Brilloin zone we have
\begin{align}
H_{M/A}(\kk)=\left(\begin{array}{cc}
H_{X}(\kk) & \Lambda_{\text{SOC}} \\ 
\Lambda^\dagger_{\text{SOC}} & H_{Y}(\kk)
\end{array} \right)\label{Eq:Hamiltonian M/A}
\end{align}
where
\begin{align}
H_{X/Y}(\kk)&=\left(\begin{array}{cc}
\frac{k^2}{2m_1}-\epsilon^{e}_{yz/xz}-\mu\mp\frac{a_1}{2}(k_x^2-k_y^2)\mp\Phi^{e}_{xz,yz} & -i\nu_{X/Y}(\kk) \\ 
i\nu_{X/Y}(\kk) & \frac{k^2}{2m_3}-\epsilon^{e}_{xy}-\mu\mp\frac{a_3}{2}(k_x^2-k_y^2)\mp\Phi_{4}+\Delta\epsilon_{xy}
\end{array} \right)\label{Eq:HamiltonianXYpoint}\\
&=A_{X/Y}\tau_0+B_{X_1,Y_1}\tau_1+B_{X_3/Y_3}\tau_3,\quad	\Lambda_{\text{SOC}}=\frac{\lambda^{e}_{\text{SOC}}}{2}\left(\begin{array}{cc}
& i \\
1 & 
\end{array}
\right)
\end{align}
corresponds to states 
with
\begin{align}
A_{X/Y}&=\frac{k^2}{2}\left(\frac{1}{2m_1}+\frac{1}{2m_3}\right)-\frac{1}{2}\left(\epsilon^{e}_{yz/xz}+\epsilon^{e}_{xy}+2\mu\right)
\mp\frac{1}{4}(a_1+a_3)(k_x^2-k_y^2)\mp \frac{\Phi^{e}_{xz,yz}+\Phi_{4}}{2}+\frac{\Delta\epsilon_{xy}}{2}\\
B_{X_3/Y_3}&=\frac{k^2}{2}\left(\frac{1}{2m_1}-\frac{1}{2m_3}\right)-\frac{1}{2}\left(\epsilon^{e}_{yz/xz}-\epsilon^{e}_{xy}\right)\mp\frac{1}{4}(a_1-a_3)(k_x^2-k_y^2)
\mp \frac{\Phi^{e}_{xz,yz}-\Phi_{4}}{2}-\frac{\Delta\epsilon_{xy}}{2}\\
B_{X_2/Y_2}&=v_{X/Y}(\kk).
\end{align}
and
\begin{align}
\nu_{X}(\kk)&=\sqrt{2}vk_y+\frac{p_1}{\sqrt{2}}\left(k_y^3+3k_yk_x^2\right)-\frac{p_2}{\sqrt{2}}k_y\left(k_x^2-k_y^2\right)\enspace,\quad \nu_{Y}(\kk)=\sqrt{2}vk_x+\frac{p_1}{\sqrt{2}}\left(k_x^3+3k_xk_y^2\right)-\frac{p_2}{\sqrt{2}}k_x\left(k_x^2-k_y^2\right).
\end{align}
In absence of SOC, diagonalization of Eq.~(\ref{Eq:HamiltonianXYpoint}) yields the band dispersion  $\xi^{\text{reg},\text{inc}}_{X/Y}=A_{X/Y}\pm\sqrt{B_{X_2/Y_2}^2+B_{X_3/Y_3}^2}$ with the "+" and "-" solution corresponding to a regular and incipient electron band at $X$ and $Y$ point, respectively. 
We define superconductivity as mean field order parameters in orbital space. The order parameters for $d_{yz}$ and $d_{xz}$  intra-orbital onsite pairing are given by
\begin{align}
\Psi^\dagger_{X,\uparrow}(\kk)\left(\begin{array}{cc}
\Delta^{e}_{yz} &  \\
& 0
\end{array}
\right)\Psi^*_{X,\downarrow}(-\kk),\quad
\Psi^\dagger_{Y,\uparrow}(\kk)\left(\begin{array}{cc}
\Delta^{e}_{xz} &  \\
& 0
\end{array}
\right)\Psi^*_{Y,\downarrow}(-\kk).
\end{align}	
Here we neglect the contribution of $d_{xy}$ orbitals to the pairing problem as discussed in the main text.
In the Nambu basis $\Psi^\dagger_{1}(\kk)=(\Psi^\dagger_{X\uparrow}(\kk),\Psi^\dagger_{Y\downarrow}(\kk),\Psi_{X\downarrow}(-\kk),-\Psi_{Y\uparrow}(-\kk))$ the BdG Hamiltonian reads $H^{M/A}_{\text{\text{\text{BdG}}}}=\sum_\kk\Psi^\dagger_1(\kk) H^{M/A}_{\text{\text{\text{BdG}}}}(\kk)\Psi_1(\kk)$, where
\begin{align}
H^{M/A}_{\text{\text{BdG}}}=\sum_\kk\Psi_1^\dagger\left(\begin{array}{cccc}
H_X & \Lambda & \hat{\Delta}_X &0  \\ 
\Lambda^\dagger & H_Y & 0 & \hat{\Delta}_Y \\ 
\hat{\Delta}_X^\dagger & 0 & -H_X & -\Lambda \\ 
0& \hat{\Delta}_Y^\dagger & -\Lambda^\dagger & -H_Y
\end{array} \right)\Psi_1\label{Eq:BdGM}
\end{align}
with the pairing matrices $\hat{\Delta}_X(\kk)=\frac{\tau_0+\tau_3}{2}\Delta^{e}_{yz}$ and $\hat{\Delta}_Y(\kk)=\frac{\tau_0+\tau_3}{2}\Delta^{e}_{xz}$.
A unitary transformation
\begin{align}
   \hat{U}^\dagger_\kk =\left(\begin{array}{cc}
	\alpha^e_{xy^X/xy^Y}(\kk) & \alpha^{e}_{xz/yz}(\kk) \\ 
	-\alpha^{e*}_{xz/yz}(\kk)& \alpha^{e*}_{xy^X/xy^Y}(\kk)
\end{array} 
\right)=
\left(\begin{array}{cc}
	\sin(\phi^{X/Y}_\kk) & i\cos(\phi^{X/Y}_\kk) \\ 
	i\cos(\phi^{X/Y}_\kk) & \sin(\phi^{X/Y}_\kk)
\end{array} 
\right)
\end{align}
transforms $H_{X/Y}$ from orbital into band space with two eigenvalues of which $\xi^{\text{reg}}_{X/Y}=A_{X/Y}+\sqrt{B_{X_2/Y_2}^2+B_{X_3/Y_3}^2}$ describe the regular bands with the corresponding eigenvectors $\ket{X_{\text{reg},\sigma}}$ and $\ket{Y_{\text{reg},\sigma}}$.
Following Ref.~\cite{Cvetkovic13} we define the projector $U_{FS}=\text{diag}(\ket{X_{\text{reg},\uparrow}},\ket{Y_{\text{reg},\downarrow}},\ket{X_{\text{reg},\downarrow}},\ket{Y_{\text{reg},\uparrow}})$ and treating SOC as a small perturbation we can transform Eq.~(\ref{Eq:BdGM}) to the reduced basis which involves states near the Fermi level only.
\begin{align}
U_{FS}^\dagger H^M_{\text{\text{BdG}}}(\kk)U_{FS}=\left(\begin{array}{cccc}
\xi^{\text{reg}}_{X}(\kk) & \frac{\lambda^{e}_{\text{SOC}}}{2}|\kappa| & \Delta^{e}_X(\kk) &  \\ 
\frac{\lambda^{e}_{\text{SOC}}}{2}|\kappa| & \xi^{\text{reg}}_{Y}(\kk) &  & \Delta^{e}_Y(\kk) \\ 
\Delta^{e\dagger}_X(\kk) &  & -\xi^{\text{reg}}_{X}(\kk) & -\frac{\lambda^{e}_{\text{SOC}}}{2}|\kappa| \\ 
& \Delta^{e\dagger}_Y(\kk) & -\frac{\lambda^{e}_{\text{SOC}}}{2}|\kappa| & -\xi^{\text{reg}}_{Y}(\kk)
\end{array} \right)\label{Eq:reducedBDG}
\end{align}
where the intra-band gaps and the orbital content are given by 
\begin{align}
    \Delta^{e}_X(\kk)&=\Delta^{e}_{yz}\cos^2(\phi^X_\kk)\\
    \Delta^{e}_Y(\kk)&=\Delta^{e}_{xz}\cos^2(\phi^Y_\kk)
\end{align}
and
\begin{align}
|a^e_{xy^X/xy^Y}(\kk)|^2&=\sin^2(\phi^{X/Y}_\kk)=\frac{1}{2}\left[1-\frac{B_{X_3/Y_3}}{\sqrt{(B_{X_2/Y_2})^2+(B_{X_3/Y_3})^2}}\right]\\
|a^e_{xz/yz}(\kk)|^2&=\cos^2(\phi^{X/Y}_\kk)=\frac{1}{2}\left[1+\frac{B_{X_3/Y_3}}{\sqrt{(B_{X_2/Y_2})^2+(B_{X_3/Y_3})^2}}\right].
\end{align}
The parameter $\frac{\lambda^{e}_\text{SOC}}{2}|\kappa|=\frac{\lambda^{e}_\text{SOC}}{2}\frac{1}{2}\left[1-\frac{B_{Y_3}B_{X_3}}{\sqrt{B_{Y_3}^2+B_{X_2}^2}\sqrt{B_{X_3}^2+B_{Y_2}^2}}\right]^{1/2}$ determines the SOC induced splitting between inner and outer electron pocket.   

By performing another unitary transformation we bring Eq.~(\ref{Eq:reducedBDG}) into band basis desribing the states at $M$ and $A$ point in presence of weak SOC.
\begin{align}
U^\dagger H^{A/M}_{\text{\text{BdG}}}(\kk)U=\left(\begin{array}{cccc}
\xi^{e}_{\text{in}}(\kk) & 0 & \Delta^{e}_\text{\text{in}}(\kk) & \Delta^{e}_\text{\text{inter}}(\kk)  \\ 
0 & \xi^{e}_{\text{out}}(\kk) &\Delta^{e}_\text{\text{inter}}(\kk)  &\Delta^{e}_\text{\text{out}}(\kk) \\ 
\Delta^{e*}_\text{\text{in}}(\kk)& \Delta^{e*}_\text{\text{inter}} & -\xi^{e}_{\text{in}}(\kk) &0  \\ 
\Delta^{e*}_\text{\text{inter}}& \Delta^{e*}_\text{\text{out}}(\kk) &0& -\xi^{e}_{\text{out}}(\kk)
\end{array} \right)\label{Eq:M point BDG}
\end{align}
The normal state band dispersion of Eq.~(\ref{Eq:M point BDG}) is given by $\xi^{e}_{\text{in/out}}(\kk)=\frac{\xi^{\text{reg}}_{X}(\kk)+\xi^{\text{reg}}_{Y}(\kk)}{2}\pm\sqrt{\frac{(\xi^{\text{reg}}_{X}(\kk)-\xi^{\text{reg}}_{Y}(\kk))^2}{4}+(\frac{\lambda^{e}_{\text{SOC}}}{2}|\kappa|)^2}$ and the superconducting intra- and inter-band order parameters can be parameterized as
\begin{align}
\Delta^{e}_\text{\text{out}}(\kk)&=\Delta^{\text{reg}}_{X}(\kk)\sin^2(\phi^e_\kk)+\Delta^{\text{reg}}_{Y}(\kk)\cos^2(\phi^e_\kk)\\
\Delta^{e}_\text{\text{in}}(\kk)&=\Delta^{\text{reg}}_{X}(\kk)\cos^2(\phi^e_\kk)+\Delta^{\text{reg}}_{Y}(\kk)\sin^2(\phi^e_\kk)\\
\Delta^{e}_\text{\text{inter}}(\kk)&=\frac{1}{2}(\Delta^{\text{reg}}_{X}(\kk)-\Delta^{\text{reg}}_{Y}(\kk))\sin(2\phi^e_\kk)\label{Eq:MInterGap}.
\end{align}
where
\begin{align}
\sin^2(\phi^e_\kk)=\frac{1}{2}\left[1-\frac{\xi^{\text{reg}}_{X}(\kk)-\xi^{\text{reg}}_{Y}(\kk)}{\sqrt{(\xi^{\text{reg}}_{X}(\kk)-\xi^{\text{reg}}_{Y}(\kk))^2+(\lambda^{e}_{\text{SOC}}|\kappa|)^2}}\right],\quad\cos^2(\phi^e_\kk)=\frac{1}{2}\left[1+\frac{\xi^{\text{reg}}_{X}(\kk)-\xi^{\text{reg}}_{Y}(\kk)}{\sqrt{(\xi^{\text{reg}}_{X}(\kk)-\xi^{\text{reg}}_{Y}(\kk))^2+(\lambda^{e}_{\text{SOC}}|\kappa|)^2}}\right].
\end{align}
For the gap at the peanut shaped pocket we find for small SOC $\Delta^{e}_\text{\text{out}}(\kk)\approx\Delta^{\text{reg}}_{X}(\kk)$.

The fitting parameters for the dispersion near the $A$ point listed in Tab.~(\ref{Tab.FittingA}).

\begin{table}[h!]
	\begin{tabular}{c |c c c c c}	
		&$M$ & &$A$  & \\ 
		\hline 
		$\epsilon^{e}_{yz/xz}$ &$30.6$& & $30.6$ &&$\text{meV}$ \\ 
		
		$\epsilon^{e}_{xy}$ &$48.6$ & & $48.6$ &&$\text{meV}$ \\ 
		
		$\frac{1}{2m_1}$ &$10.2060$& & $4.54$ &&$\text{meV}\enspace\mathring{A}^2$ \\ 
		
		$\frac{1}{2m_3}$ &$1355.9$& & $602.64$ &&$\text{meV}\enspace\mathring{A}^2$ \\ 
		
		$\alpha_1$ &$991.44$& & $440.64$ &&$\text{meV}\mathring{A}^2$ \\ 
		
		$\alpha_3$ &$-2937.9$& & $-1305.7$ &&$\text{meV}\enspace\mathring{A}^2$ \\ 
		
		$v$ &$-329.4$& & $-219.6$ &&$\text{meV}\enspace\mathring{A}$ \\ 
		
		$p_{z_1}$ &$-2700.9$& & $-800.27$ &&$\text{meV}\enspace\mathring{A}^3$ \\
		
		$p_{z_2}$ &$-229.7$&  & $-68.06$ && $\text{meV}\enspace\mathring{A}^3$\\
		
		$\lambda^{e}_{\text{SOC}}$ &$4$& & $4$ &&$\text{meV}$ \\	
		
		$\Phi^{e}_{xz,yz}$ &$-26$& & $-26$ &&$\text{meV}$ \\
		
		$\Phi_{4}$ &$45$& & $45$ && $\text{meV}$\\	
       $\Delta\epsilon_{xy}$ &$40$ & & $40$ && $\text{meV}$	
	\end{tabular} 
	
	\caption{Fitting parameters for the states at the $M$ and $A$ point fitted to ARPES data at 10 K.\label{Tab.FittingA} }
\end{table}

\begin{table}[h!]
		\begin{center}
			\label{tab:table4}
			\begin{tabular}{c|c|c|c|c}
			\textbf{Pocket} & \multicolumn{2}{c}{$\mathbf{k_{Fx}} / \AA^{-1}$} & \multicolumn{2}{c}{$\mathbf{k_{Fy}} / \AA^{-1}$} \\ 
			\hline
			& \textbf{Model} & \textbf{Exp} & \textbf{Model} & \textbf{Exp} \\
			\hline
				$\Gamma$ & 0.042 & 0.033 \cite{Watson2015} &0.063 &0.077 \cite{Watson2015}\\
				$M$ & 0.138 &0.14 \cite{Watson2015}& 0.032 &0.02 \cite{Watson2015}\\
				$Z$ & 0.092 &0.10 \cite{Watson2017b}& 0.1407  &0.15 \cite{Watson2017b}\\
				$A$ & 0.206 &0.19 \cite{Watson2017b}& 0.047 &0.03 \cite{Watson2017b}\\
			\end{tabular}
			\caption{Comparison of simulated $k_F$ values from the orbital projective model, with available ARPES data.}
		\end{center}
	\end{table}

\section*{Supplementary note 3}
In this note, we discuss \textbf{the pairing interaction}.
We assume superconductivity to be driven by repulsive onsite interactions involving only intra-orbital interactions that translate mostly into pure inter-band pair-hopping interaction between hole and electron pockets and between both electron pockets corresponding to $(\pi,0),(0,\pi)$ and $(\pi,\pi)$ spin-fluctuation mediated pairing, respectively. Moreover, we neglect interaction within $d_{xy}$ orbitals. The Hamiltonian was introduced in Ref.~\cite{Kang_s+id_2018} and is given by
\begin{align}
H_{\text{int}}=&U_{\text{he}}\sum_{\kk,\kk^\prime,\mu}d^\dagger_{\mu,\kk,\uparrow}d^\dagger_{\mu,-\kk,\downarrow}d_{\mu,-\kk^\prime+\Qv_\mu,\downarrow}d_{\mu,\kk^\prime+\Qv_\mu,\uparrow}\nn\\
+&J'_{\text{he}}\sum_{\kk,\kk^\prime,\mu\neq\nu}d^\dagger_{\mu,\kk,\uparrow}d^\dagger_{\mu,-\kk,\downarrow}d_{\nu,-\kk^\prime+\Qv_\nu,\downarrow}d_{\nu,\kk^\prime+\Qv_\nu,\uparrow}\nn\\
+&J'_{\text{ee}}\sum_{\kk,\kk^\prime,\mu\neq\nu}d^\dagger_{\mu,\kk+\Qv_\mu,\uparrow}d^\dagger_{\mu,-\kk+\Qv_\mu,\downarrow}d_{\nu,-\kk^\prime+\Qv_\nu,\downarrow}d_{\nu,\kk^\prime+\Qv_\nu,\uparrow}+h.c\label{Eq:InteractionHamiltonianAppendix}
\end{align}
We treat Eq.~(\ref{Eq:InteractionHamiltonianAppendix}) using a mean-field decomposition into the pairing terms introduced above:
\begin{align}
\Psi^\dagger_{\Gamma/Z,\uparrow}(\kk)\left(
\begin{array}{cc}
\Delta^{h}_{xz} &  \\
& \Delta^{h}_{yz}
\end{array}
\right)\Psi^*_{\Gamma/Z,\downarrow}(-\kk)\quad,\quad
\Psi^\dagger_{M/A,\uparrow}(\kk)\left(\begin{array}{cccc}
\Delta^{e}_{yz} &  &  &  \\
&0  &  &  \\
&  & \Delta^{e}_{xz} &  \\
&  &  & 0
\end{array}
\right)\Psi^*_{M/A,\downarrow}(-\kk)\label{Eq:OrderParameters}
\end{align}	
From now on we neglect contributions from inter-band pairing gaps (such as Eq.~(\ref{Eq:GammaInterGap}) and (\ref{Eq:MInterGap})) and write the free energy as
\begin{align}
F=-\hat{\Delta}^\dagger \hat{U}^{-1}\hat{\Delta}-\sum_{\kk,\alpha} \left[E_\alpha(\kk)+2T\ln\left(1+\exp\left(-\frac{E_\alpha(\kk)}{T}\right)\right)\right]+\sum_{\kk,\alpha}\xi_{\alpha}(\kk)\label{SaddlePoint}
\end{align}
where $\alpha\in\{h_\text{in},h_\text{out},e_\text{in},e_\text{out},e_{\text{inc}_1},e_{\text{inc}_2}\}$ labels the bands in presence of SOC and $\hat{\Delta}=\left(\Delta^{h}_{xz},\Delta^{h}_{yz},\Delta^{e}_{yz},\Delta^{e}_{xz}\right)^T$ the gaps in orbital space and $E_\alpha(\kk)=\sqrt{\xi_{\alpha}(\kk)^2+|\Delta_\alpha(\kk)|^2}$. The pairing matrix is given by
\begin{align}
\hat
{U}=\left(\begin{array}{cccc}
0 & 0 & J^\prime_{\text{he}} & U_{\text{he}} \\ 
0 & 0 & U_{\text{he}} & J^\prime_{\text{he}} \\ 
J^\prime_{\text{he}} &U_{\text{he}} & 0 & J'_{\text{ee}} \\ 
U_{\text{he}} & J^\prime_{\text{he}} & J'_{\text{ee}} & 0
\end{array} \right).
\end{align}
Minimizing the free energy with respect to $\hat{\Delta}^\dagger$ and focusing on the larger hole pocket and the two electron bands near the Fermi surface (i.e $\{h_\text{in},e_\text{in},e_\text{out}\}$) yields the BCS gap-equations

\begin{align}
\Delta^{h}_{xz}=-\sum_{\kk}\Bigg[&J^\prime_{\text{he}}\left[\frac{\cos^2(\phi^X_\kk)\cos^2(\phi^e_\kk)\Delta^{e}_{\text{in}}(\kk)}{2E^e_\text{in}(\kk)}\tanh\left(\frac{E^e_\text{in}}{2T}\right)+\frac{\cos^2(\phi^X_\kk)\sin^2(\phi^e_\kk)\Delta^{e}_{\text{out}}(\kk)}{2E^e_\text{out}(\kk)}\tanh\left(\frac{E^e_\text{out}}{2T}\right)\right]\nn\\
+&U_{\text{he}}\left[\frac{\cos^2(\phi^Y_\kk)\sin^2(\phi^e_\kk)\Delta^{e}_{\text{in}}(\kk)}{2E^e_\text{in}}\tanh\left(\frac{E^e_\text{in}}{2T}\right)+\frac{\cos^2(\phi^Y_\kk)\cos^2(\phi^e_\kk)\Delta^{e}_{\text{out}}(\kk)}{2E^e_\text{out}(\kk)}\tanh\left(\frac{E^e_\text{out}}{2T}\right)\right]\Bigg]\nn\\
\Delta^{h}_{yz}=-\sum_{\kk}\Bigg[&U_{\text{he}}\left[\frac{\cos^2(\phi^X_\kk)\cos^2(\phi^e_\kk)\Delta^{e}_{\text{in}}(\kk)}{2E^e_\text{in}}\tanh\left(\frac{E^e_\text{in}}{2T}\right)+\frac{\cos^2(\phi^X_\kk)\sin^2(\phi^e_\kk)\Delta^{e}_{\text{out}}(\kk)}{2E^e_\text{out}(\kk)}\tanh\left(\frac{E^e_\text{out}}{2T}\right)\right]\nn\\
+&J^\prime_{\text{he}}\left[\frac{\cos^2(\phi^Y_\kk)\sin^2(\phi^e_\kk)\Delta^{e}_{\text{in}}(\kk)}{2E^e_\text{in}}\tanh\left(\frac{E^e_\text{in}}{2T}\right)+\frac{\cos^2(\phi^Y_\kk)\cos^2(\phi^e_\kk)\Delta^{e}_{\text{out}}(\kk)}{2E^e_\text{out}(\kk)}\tanh\left(\frac{E^e_\text{out}}{2T}\right)\right]\Bigg]\nn\\
\Delta^{e}_{yz}=-\sum_{\kk}\Bigg[&J^\prime_{\text{he}}\frac{\sin^2(\phi^h_\kk)\Delta^{h}_{\text{out}}(\kk)}{2E^h_\text{out}(\kk)}\tanh\left(\frac{E^h_\text{out}}{2T}\right)
+U_{\text{he}}\frac{\cos^2(\phi^h_\kk)\Delta^{h}_{\text{out}}(\kk)}{2E^h_\text{out}(\kk)}\tanh\left(\frac{E^h_\text{out}}{2T}\right)\nn\\
+&J^\prime_{\text{ee}}\left[\frac{\cos^2(\phi^Y_\kk)\sin^2(\phi^e_\kk)\Delta^{e}_\text{in}(\kk)}{2E^e_\text{in}(\kk)}\tanh\left(\frac{E^e_\text{in}}{2T}\right)+\frac{\cos^2(\phi^Y_\kk)\cos^2(\phi^e_\kk)\Delta^{e}_\text{out}(\kk)}{2E^e_\text{out}(\kk)}\tanh\left(\frac{E^e_\text{out}}{2T}\right)\right]\Bigg]\nn\\
\Delta^{e}_{xz}=-\sum_{\kk}\Bigg[&U_{\text{he}}\frac{\sin^2(\phi^h_\kk)\Delta^{h}_{\text{out}}(\kk)}{2E^h_\text{out}(\kk)}\tanh\left(\frac{E^h_\text{out}}{2T}\right)
+J^\prime_{\text{he}}\frac{\cos^2(\phi^h_\kk)\Delta^{h}_{\text{out}}(\kk)}{2E^h_\text{out}(\kk)}\tanh\left(\frac{E^h_\text{out}}{2T}\right)\nn\\
+&J'_{\text{ee}}\left[\frac{\cos^2(\phi^X_\kk)\cos^2(\phi^e_\kk)\Delta^{e}_\text{in}(\kk)}{2E^e_\text{in}(\kk)}\tanh\left(\frac{E^e_\text{in}}{2T}\right)+\frac{\cos^2(\phi^X_\kk)\sin^2(\phi^e_\kk)\Delta^{e}_\text{out}(\kk)}{2E^e_\text{out}(\kk)}\tanh\left(\frac{E^e_\text{out}}{2T}\right)\right]\Bigg].\nn\\\label{Eq:GEq}
\end{align}
Note that the system of equations (\ref{Eq:GEq}) is not a closed one, as the chemical potential $\mu$  depends on the total number of particles (N) which we fix  at 10 K in the non-superconducting orthohombic state as
\begin{align}
N(\Delta(T),\mu(T))=N(0,\mu(10K))=-\frac{\partial F}{\partial \mu}=\sum_{\alpha,\kk}1-\frac{\xi_\alpha(\kk)}{E_\alpha(\kk)}\tanh\left(\frac{E_\alpha(\kk)}{2T}\right)\label{Eq:NumberEquation}.
\end{align}
A self-consistent treatment of the pairing problem requires solving the coupled set of equations (\ref{Eq:GEq}) and (\ref{Eq:NumberEquation}).

\section*{Supplementary note 4}
In this note, we discuss the \textbf{10-orbital tight binding model} used in part D of the results section in the main text.

	\begin{figure}[t]
	\centering
	\includegraphics[width = \linewidth]{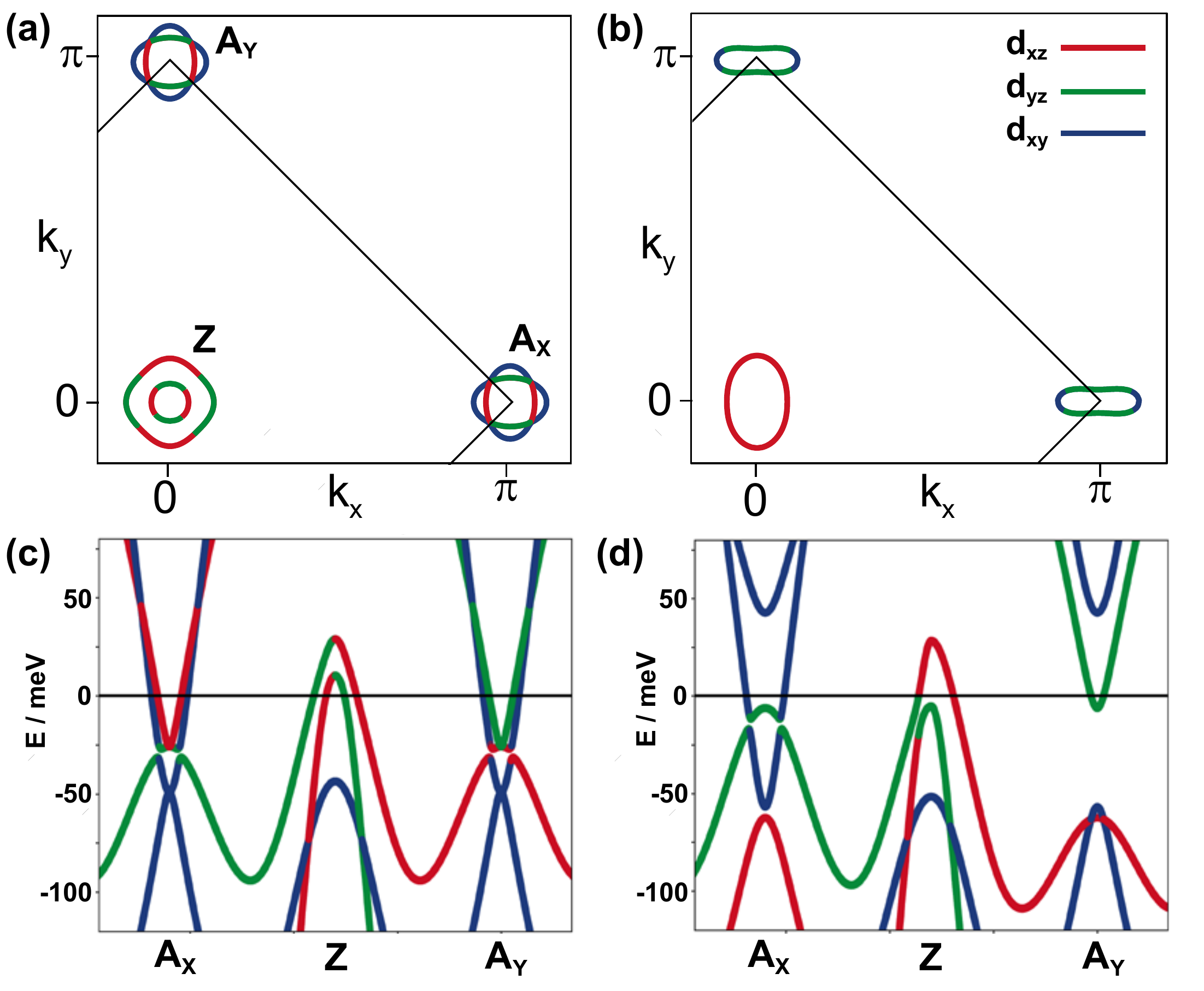}
	\caption{Electronic structure of the tetragonal and nematic state of FeSe calculated from a tight binding model fit to ARPES data. a) Fermi surface at $k_z = \pi$ for the tetragonal state, the black solid line describes the Brillouin zone boundary of the two atom unit cell. b) Equivalent Fermi surface with the inclusion of the nematic order parameter. c,d) Band dispersions along the path $A_x$-$Z$-$A_y$, which corresponds to $(\pi,0,\pi) - (0,0,\pi) - (0,\pi,\pi)$ in $k$-space, for the tetragonal and nematic state respectively.}
	\label{fig:fig1_LR}
\end{figure}

The tight binding model uses the form and notation of Ref. \cite{Eschrig2009} and the hopping parameters, in units of eV, are

\underline{\textbf{2D parameters}}
\begin{align}
t^{11}_{11}  &=  0.01818 
&& t^{10}_{16}  = -0.03133
&&t^{11}_{33}  =  0.02433 
\nonumber \\
t^{20}_{11}  &=  0.00093 
&&t^{21}_{16}  = -0.00231
&&t^{20}_{33}  = 0.00096 
\nonumber \\
t^{11}_{13}  &=-0.01226i
&&t^{10}_{18}  =  0.11516i
&&t^{02}_{33}  = -0.00717 
\nonumber \\
t^{11}_{15}  &= -0.01817
&&t^{10}_{27}  = -0.04988
&&t^{22}_{33}  =  0.00758 
\nonumber \\
t^{11}_{22}  &= -0.01669 
&&t^{10}_{29}  = -0.09492i
&&t^{10}_{38}  =  0.00868
\nonumber \\
t^{11}_{23}  &=  0.01484i 
&&t^{10}_{2,10} =  0.059659
&&t^{21}_{38}  = -0.00493
\nonumber \\
t^{11}_{34}  &=  0.01650 
&&t^{11}_{35}  =  0.00569i 
&&t^{10}_{4,10} =  -0.00902i
\nonumber \\
t^{10}_{49}  &=  0.05023
&&t^{21}_{49} = -0.00008
\nonumber \\
\epsilon_1     &=  0.03405
&&\epsilon_2     = 0.0200 
&&\epsilon_3     =  0.00310
\nonumber \\
\epsilon_4    &=  0.00310 
&&\epsilon_5     = -0.19398 \nonumber
\end{align}
\underline{\textbf{3D parameters}}
\begin{align}
t^{101}_{16} &= 0.00270
&&t^{001}_{11} = 0
&&t^{201}_{14} = 0.00950i
\nonumber \\
t^{121}_{16} &= -0.00283
&&t^{111}_{11} = 0
&&t^{101}_{19} = 0.00333i
\nonumber \\
t^{101}_{18} &= 0.00150i
&&t^{201}_{11} =  0.00013
&&t^{121}_{19} = 0.00517i
\nonumber \\
t^{001}_{33} &= 0.00183
&&t^{101}_{38} = 0.00300
&&t^{121}_{49} = 0.00100.
\nonumber \\
t^{201}_{33} &= -0.00133
&&t^{121}_{38} = -0.00050
&&t^{101}_{49} = 0.00217
\nonumber \\
t^{021}_{33} &= 0.00333
&&t^{101}_{39} = 0.00250 
\nonumber
\end{align}

We additionally include spin orbit coupling in the standard form \cite{Saito2015},

\begin{equation}
\hat{H}_\text{SOC}(k) = 
\begin{pmatrix}
H(\mathbf{k}) + \hat{L}_z & \hat{L}_x + i\hat{L}_y  \\
\hat{L}_x - i\hat{L}_y & H(\mathbf{k}) - \hat{L}_z
\end{pmatrix}.
\end{equation}

\noindent Where $\hat{L}_i$ is defined as 

\begin{equation}
\hat{L}_i = \frac{\lambda_i}{2}
\begin{pmatrix}
\hat{l}_i & 0  \\
0 & \hat{l}_i
\end{pmatrix}.
\end{equation}

\begin{centering}
\begin{equation}
\hat{l}_x =
\begin{pmatrix}
0 & 0 & -i & 0 & 0 \\
0 & 0 & 0 & i & 0 \\
i & 0 & 0 & 0 & 0 \\
0 & -i & 0 & 0 &  -\sqrt{3} i \\
0 & 0 & 0 &  \sqrt{3} i & 0 
\end{pmatrix}
\end{equation}
\begin{equation}
\hat{l}_y =
\begin{pmatrix}
0 & 0 & 0 & i & 0 \\
0 & 0 & i & 0 & 0 \\
0 & -i & 0 & 0 & \sqrt{3} i \\
-i & 0 & 0 & 0 & 0\\
0 & 0 & -\sqrt{3} i  &  & 0 
\end{pmatrix}
\end{equation}

\begin{equation}
\hat{l}_z =
\begin{pmatrix}
0 & 2i & 0 & 0 & 0 \\
-2i & 0 & 0 & 0 & 0 \\
0 & 0 & 0 & -i & 0 \\
0 & 0 & i & 0 &  0\\
0 & 0 & 0 &  0 & 0 
\end{pmatrix}
\end{equation}
\end{centering}

\noindent$\hat{l}_x$, $\hat{l}_y$ and $\hat{l}_z$ describe the angular momentum operators in the $x$,$y$, and $z$ directions respectively for a single Fe atom where the rows and columns are defined equivalent to the Hamiltonian as $[d_{xy},d_{x^2-y^2},d_{xz},d_{yz},d_{z^2}]$. 
In this work we set the spin orbit coupling strength such that $\lambda_z =19$~meV, and $\lambda_{x/y} = 5$~meV \cite{Rhodes2019}. The band structure and Fermi surface with and without the nematic order parameter defined in the main text are presented in FIG. SM1.

\section*{Supplementary note 5}
In this note, we discuss the mathematical form of \textbf{the pairing vertex for the linearised superconducting gap equation} of Eq.~(13) in the main text for the 10-orbital tight binding model.

The pairing vertex for spin fluctuation mediated superconductivity in the presence of spin orbit coupling is defined \cite{Saito2015,Rhodes2018} as
 \begin{equation}
\Gamma^\text{SOC}_{\mu\nu}(\mathbf{k},\mathbf{k'}) = \Big[\Gamma^{\Uparrow\Downarrow\Uparrow\Downarrow}_{\mu\nu}(\mathbf{k},\mathbf{k'}) - \Gamma^{\Uparrow\Downarrow\Downarrow\Uparrow}_{\mu\nu}(\mathbf{k},\mathbf{k'})\Big],
\end{equation}

 where
\begin{equation}
\Gamma^{\Sigma\bar{\Sigma}\Lambda\bar{\Lambda}}_{\mu\nu}(\mathbf{k},\mathbf{k'}) =  \sum_{stpq}\sum_{\sigma\bar{\sigma}\lambda\bar{\lambda}} a_{\mu\Sigma}^{t\sigma*}(\mathbf{k}) a_{\mu\bar{\Sigma}}^{s\bar{\sigma}*}(-\mathbf{k}) Re[\Gamma^{pq;\lambda\bar{\lambda}}_{st;\sigma\bar{\sigma}}(\mathbf{k},\mathbf{k'})] a_{\nu\bar{\Lambda}}^{p\bar{\lambda}}(\mathbf{-k'}) a_{\nu\Lambda}^{q\lambda}(\mathbf{k}').
\end{equation}
Here $a_{\mu\Sigma}^{t\sigma}(\mathbf{k})$ is the eigenvector of the original Hamiltonian in the presence of spin orbit coupling which connects the orbital  and spin basis ($s,p,q,t$ and $\sigma,\lambda$) with the band and pseudospin - band basis ($\mu,\nu$ and $\Sigma,\Lambda$). 

The pairing vertex in orbital space \cite{Saito2015} is then defined as 
\begin{equation}
\Gamma^{pq;\lambda\bar{\lambda}}_{st;\sigma\bar{\sigma}}(\mathbf{k},\mathbf{k'})  = V^{c}_{pq;st}\delta_{\sigma\lambda}\delta_{\bar{\sigma}\bar{\lambda}} + V^{s}_{pq;st}\mathbf{\vec{\sigma}}_{\sigma\lambda}\mathbf{\vec{\sigma}}_{\bar{\sigma}\bar{\lambda}}.
\end{equation}
\noindent Here, $\mathbf{\vec{\sigma}}_{\sigma\lambda}$ is a vector of Pauli matrices. By performing the spin sum we obtain a conditional pairing vertex
\begin{equation}
\Gamma^{pq;\lambda\bar{\lambda}}_{st;\sigma\bar{\sigma}}(\mathbf{k},\mathbf{k'})]  =
\begin{cases}
V^{c}_{pq;st} + V^{s}_{pq;st}, & \sigma = \lambda = \bar{\sigma} = \bar{\lambda} \\
V^{c}_{pq;st} - V^{s}_{pq;st}, & \sigma = \lambda \neq \bar{\sigma} = \bar{\lambda} \\
2V^{s}_{pq;st}, & \sigma = \bar{\lambda} \neq \lambda = \bar{\sigma}  \\
0, & otherwise.
\end{cases}
\end{equation}

Where $V^{c/s} = \frac{1}{2}U^{c/s}\chi^{c/s}U^{c/s}$, $\chi^{c/s} = \chi^0[1\pm U^{c/s}\chi^0]^{-1}$ is the RPA enhanced spin susceptibility in the charge (c) or spin (s) channel and 

\begin{equation}
U^{s} =
\begin{cases}
U, & p= q = s = t \\
U', & p = s \neq q = t \\
J, & p = q \neq s = t \\
J', & p = t \neq q = s \\
0, & otherwise,
\end{cases}
\end{equation}

\begin{equation}
U^{c} =
\begin{cases}
U, & p = q = s = t \\
-U'+2J, & p = s \neq q = t \\
2U'-J, & p = q\neq s = t \\
J', & p = t \neq q = s \\
0, & otherwise.
\end{cases}
\end{equation}
In these calculations we assume spin rotational invariance, $U' = U - 2J$, $J = \frac{U}{6}$ and $J' = J$. U is set to 0.5~eV.

\end{widetext}
\end{document}